\documentclass[12pt,a4paper]{article}
\pdfoutput=1

\usepackage[utf8]{inputenc}
\usepackage[T1]{fontenc}
\usepackage[margin=2.0cm]{geometry}
\usepackage[pdfstartview={FitH},bookmarks=false,linktoc=page,colorlinks=true,linkcolor=blue,citecolor=blue,urlcolor=blue]{hyperref}
\usepackage[numbered]{bookmark}
\usepackage{mathtools}
\usepackage{amssymb}
\usepackage{graphicx}
\usepackage[font=small,labelfont=bf]{caption}
\usepackage{authblk}
\usepackage{cite}
\usepackage{dsfont}
\usepackage{xcolor}
\usepackage[titles]{tocloft}
\usepackage{float}
\usepackage{subcaption}
\usepackage{color}
\usepackage{tikz}
\usepackage{ifthen}
\usepackage[warn]{textcomp}
\numberwithin{equation}{section}
\allowdisplaybreaks
\usepackage{multirow}

\usetikzlibrary{patterns}

\newcommand{\ket}[1]{|#1\rangle}

\newcommand{\braket}[1]{\langle#1 \rangle}

\begin{document}
	
	\title{\vspace{2cm}\textbf{Holographic Fisher Information Metric in Schr\"odinger Spacetime}\vspace{1cm}}
	\author[a,b]{H. Dimov}
	\author[a]{I. N. Iliev}
	\author[a]{M. Radomirov}
	\author[a,c]{R. C. Rashkov}
	\author[a,b]{T. Vetsov}
	
	\affil[a]{\textit{Department of Physics, Sofia University,}\authorcr\textit{5 J. Bourchier Blvd., 1164 Sofia, Bulgaria}
		
		\vspace{-10pt}\texttt{}\vspace{0.0cm}}
	
  \affil[b]{\textit{The Bogoliubov Laboratory of Theoretical Physics,
    JINR,}\authorcr\textit{141980 Dubna, Moscow region, Russia}
		
    \vspace{-10pt}\texttt{}\vspace{0.0cm}}
	
  \affil[c]{\textit{ Institute for Theoretical Physics, Vienna University of
    Technology,}\authorcr\textit{Wiedner Hauptstr. 8–10, 1040 Vienna, Austria}
		
    \vspace{10pt}\texttt{h\_dimov,ivo.iliev,radomirov,rash,vetsov@phys.uni-sofia.bg}\vspace{0.1cm}}
    \date{} \maketitle
	
  \begin{abstract} In this paper we study the Fisher information metric on the
    space of the coupling constants on both sides of the duality between
    non-relativistic dipole field theories and string theory in Schr\"odinger
    spacetime. We consider the following setup. In the gauge theory side one
    can deform a given conformal field theory by a proper scalar operator and
    compute the quantum information metric via the two-point correlation
    function between two such operators. On the string side the deformation
    corresponds to a scalar field probing the background. In the large $N$
    limit of the theory the probing can be done without backreaction on the
    original spacetime, thus one can construct a perturbative scheme for the
    calculation of the dual holographic Fisher information metric as shown by
    \cite{Trivella:2016brw}. Considering the asymptotic behaviour of the holographic Fisher information metric close to the boundary of the Schr\"odinger spacetime we show that its divergence structure exactly matches its dual quantum counterpart up to the leading order, thus extending the holographic setup up to the non-relativistic case. One should note that the existence of other terms are not seen from the boundary theory to this level of approximation. Their behaviour near the boundary however, is pointing what kind of information from the boundary theory is missing to be able to reconstruct the bulk. Obviously more work is needed to refine and elucidate their meaning and interrelations in holographic setup.   \end{abstract}
	
  \thispagestyle{empty}
	\newpage
  \tableofcontents
	
  \section{Introduction}
	
  The advancement of string theory and the discovery of the AdS/CFT
  correspondence \cite{Maldacena:1997re} have brought us a great deal of
  understanding about the nature of different high-energy physics models and
  their intricate interrelations. One particularly useful property of this
  correspondence is that it can relate perturbatively computable
  characteristics of a higher dimensional string theory to the degrees of
  freedom of lower dimensional strongly correlated quantum field theory.
  This is also valid in the other direction, when the quantum system is weakly coupled
  and its dual gravitational counterpart is at strong coupling. In this
  context, the string/gauge duality opens a window to study non-perturbative
  phenomena with well-known analytical techniques.
	
  Recent attempts to generalize the AdS/CFT correspondence to strongly coupled
  non-relati-vistic field theories \cite{Son:2008ye,Balasubramanian:2008dm}
  have lead to the construction of various classes of background solutions.
  Particularly interesting examples include the non-relativistic Schr\"odinger
  spacetimes, where the isometry group of the solutions on the string side is
  the Schr\"odinger group. It consists of time and space translations, space
  rotations, Galilean boosts, dilatations and special conformal
  transformations. The quantum duals to such models fall in the class of the so
  called dipole gauge theories, which are characterized with non-locality (see
  for instance \cite{Bergman:2000cw,Iso:2000ew, Bergman:2001rw}). For
  a detailed group-theoretical perspective on non-relativistic holography see
  \cite{Dobrev:2013kha}.
	
  An important understanding of non-relativistic holography has been revealed
  in \cite{Guica:2017mtd}, where strong arguments for integrability and
  quantitative matching between string and gauge theory predictions have been
  presented. These studies have lead to a number of interesting applications of
  non-relativistic holography in condensed matter physics and string theory
  such as the description of ordinary $\mathcal{N}$ = 1 SQCD-like gauge
  theories considered in the context of D-brane constructions
  \cite{Gursoy:2005cn, Freedman:2005cg, Gursoy:2006gm,Chu:2006ae,Bobev:2006fg,
Bobev:2007bm, Bobev:2005cz}, the Sachdev-Ye-Kitaev (SYK) model
  \cite{Sachdev:1992fk}, Fermi unitary gas \cite{Shin2008}, and models with
  trapped supercooled atoms \cite{Son:2008ye, Adams:2008wt}, which in most
  cases are strongly correlated. For these reasons, and the fact that currently
  very little is known for dipole theories, we are motivated to investigate the
  properties of such holographic models on both sides of the correspondence.
  Further studies of non-relativistic holography in Schr\"odinger spacetime can
  be found in \cite{Golubtsova:2020mjn, Golubtsova:2020fpm, Dimov:2019koi,
  Zoakos:2020gyb, Georgiou:2020qnh, Georgiou:2019lqh, Georgiou:2018zkt,
Ahn:2017bio}.
	
Recently, an interesting subject in holography gained popularity after the
conjecture of Ryu-Takayanagi\cite{Ryu:2006bv, Ryu:2006ef}, suggesting
  a holographic relation between quantum entanglement and codimension two
  extremal surfaces in the dual bulk gravitational theory.  Consequent studies
  of complexity and related concepts \cite{Susskind:2014rva, Stanford:2014jda}
  brought further interest to the information-theoretic analysis of holographic
systems. One of the key constructions in these investigations turns out to be the
  quantum Fisher information metric (QFIM), which plays an important role not
  only in quantum information theory, but also in high-energy physics. As advocated in \cite{MIyaji:2015mia} and subsequent works, the CFT QFIM is approximately dual to the volume of a codimension one time slice in AdS space (see however \cite{Moosa:2018mik, Belin:2018bpg} for a recent critique of this proposal), which on the other hand was also conjectured in \cite{Susskind:2014rva, Stanford:2014jda} to give a measure of complexity of the system under consideration.
  
  Further fruitful applications of QFIM and its
  holographic dual metric include phenomena and models such as quantum
  information scrambling \cite{Sharma:2020eez, Touil:2020vuz,
  PhysRevX.9.031048}, quantum metrology \cite{Liu:2019xfr}, canonical
  energy-momentum tensor \cite{Lashkari:2015hha}, quantum phase transitions
  \cite{Dimov:2017ryz, vetsov2018}, entanglement entropy \cite{ Dimov:2017ryz,
  Dimov:2017sfk,Banerjee:2017pt,Sarkar:2017pjp}, bulk reconstruction
  \cite{VanRaamsdonk:2010pw, Tsuchiya:2020drh}, instantons \cite{Malek:2015hea}
  and many others (see for instance \cite{Erdmenger:2020vmo,
  Trivella:2019fbe,Karar:2019bwy}). Additionally, whenever it is possible
  to extrapolate QFIM to certain thermodynamic limits, one can use it to
  describe the distance between classically measurable statistical macrostates
  with various applications in condensed matter physics \cite{Ruppeiner:1995zz}
  and black hole physics (see for example \cite{Mansoori:2014oia,
  HosseiniMansoori:2019jcs, Kolev:2019tqa, Dimov:2019fxp, Khan:2020kfu}). This
  further motivates us to extend the information-theoretic analysis of the
  AdS/CFT correspondence to the non-relativistic holography in Schr\"odinger
  spacetimes.
	
  In general, the concept of ``information metric'' is based on the purely
  geometric idea that one can construct a well defined Riemannian distance
  (metric) \cite{amari2007methods, amari2016information} between various micro-
  or macrostates of the system. Due to the fact that geometry studies
the mutual
  relations between elements, such as distance and curvature, one can naturally
  uncover essential features and gain valuable insights of the system under
  consideration. This is strongly evident in holography, where classical
  space-time geometry has the capacity to encode important properties of the
  dual quantum system.
	
  In this paper, we show that within the context of non-relativistic holography in
  Schr\"odinger spacetime, the Fisher information metric (FIM) on
  the space of coupling constants of the model can be explicitly calculated on
  both sides of the correspondence. This will allow us to make 
  quantitative
  and qualitative checks of the duality.  Our investigation is based on similar
  studies, conducted in \cite{Trivella:2016brw} and  \cite{Bak:2017rpp}.
  a perturbative scheme for computing the holographic Fisher information metric
  (HFIM) on the string side has been presented for AdS space. We also show that
  in certain limits HFIM in Schr\"odinger background fully reproduces the QFIM
  on the lower dimensional boundary of the spacetime, thus confirming the
  duality in this case.
	
  The structure of this paper is as follows. In Section \ref{sec2} we
  present the computational setup for the QFIM and its dual bulk
  counterpart, HFIM. In Section \ref{sec3} we compute QFIM on the dipole gauge theory side
  via the 2-point correlation function between primary operators deforming the
  original conformal field theory. In Section \ref{sec4} we use a perturbative
  method to explicitly calculate the dual HFIM in the bulk of the Schr\"odinger
  spacetime. In Section \ref{sec5} we show how the leading divergence structure of HFIM on the gravity side reduces to QFIM on the
  lower dimensional boundary of the Schr\"odinger spacetime. In Section
  \ref{sec6} we analyse the divergence structure of HFIM along the holographic
  direction. In Section \ref{sec7} we give a short review of our results. At
  the end of the paper we present some Appendices with detailed calculations of
  the QFIM and HFIM integrals, where novel results for the incomplete gamma
  functions have been obtained. 
	
  \section{Computational setup}\label{sec2}
	
  In this section we briefly discuss the computational techniques used to
  calculate the Fisher information metric on both sides of the correspondence.
  We mainly follow the presentation of \cite{Trivella:2016brw} and
  \cite{Alvarez-Jimenez:2017gus}, where a general CFT on an Euclidean
  $\mathbb{R}^D=\mathbb{R}^d\times\mathbb{R}^1$ space is considered.
	
  \subsection{CFT construction} \label{sec2-1} Let us start by considering an
  initially undeformed conformal field theory (CFT), living on an Euclidean
  $\mathbb{R}^D=\mathbb{R}^d\times\mathbb{R}^1$ space and described by an
  Euclidean Lagrangian $\mathcal{L}_0 $, defined for Euclidean time\footnote{We
  will refer to $t$ as the real time and $\tau=-it$ as the Wick-rotated time.}
  $\tau\in (-\infty, 0)$. Consequently, we perturb $\mathcal{L}_0$ at $\tau=0$
  by some quantum operators $\mathcal{O}_a(\tau)$, $a=1,\dots, n$, to a new
  theory with $\mathcal{L}_1$ for $\tau>0$, i.e.
  $\mathcal{L}_1=\mathcal{L}_0+\delta\lambda^a\mathcal{O}_a$, where
  $\delta\lambda^a$ are given real coupling constants\footnote{In general, we
  consider the situation where the original Lagrangian $\mathcal{L}_0$ has an
explicit dependence on a number of physical parameters $\lambda^a$, $a=1,\dots,
n$, thus one can deform the theory by multiple operators with corresponding
real coupling constants $\lambda^a\to\lambda^a+\delta\lambda^a$. However, it
will be qualitatively informative to consider only one such deformation.}. By
definition, the quantum information metric\footnote{ In general, the complex quantity $G_{ab}$ is called
the quantum geometric tensor. Its real part is the quantum information metric
and its imaginary part is known as the Berry curvature. Usually, the quantum
Fisher information metric is defined to be 4 times the quantum information
metric, $F_{ab}=4\Re{(G_{ab})}$. However,
with a slight abuse of notation, we will follow the terminology
established by \cite{Trivella:2016brw} and call $G_{ab}$ the quantum information metric or QIM.} (QIM) $G_{ab}$, between the ground state
$|\psi_0\rangle$ of the undeformed CFT and the ground state $|\psi_1\rangle$ of
the deformed theory, can be obtained by expanding 
the quantum fidelity\footnote{ Fidelity is a quantum measure, which specifies
the degree of change occurring in the system by turning on the deformations.
Let us note that there is also another definition, 
$\mathcal{F}'=|\langle\psi|\xi\rangle|=\sqrt \mathcal{F}$, which is
sometimes
referred to as quantum fidelity as well. However, $\mathcal{F}$ is more common,
while $\mathcal{F}'$ could be referred to as the square root fidelity.} at
temporal infinities in powers of $\delta\lambda$ 
  \begin{equation}\label{eqBala}
    \mathcal{F}(\lambda^a,\lambda^a+\delta\lambda^a)\equiv|\langle \psi_1
    (\tau\to\infty,\lambda^a+\delta\lambda^a)|\psi_0(\tau\to-\infty,\lambda^a)\rangle|^2=1-\sum\limits_{a,b=1}^n
    G_{ab}\delta\lambda^a\delta\lambda^b+\mathcal{O}(\delta\lambda^3),
  \end{equation}
  where one can show that \cite{Trivella:2016brw, Alvarez-Jimenez:2017gus}
  \begin{equation}\label{eqQIMCFTsidea} {G_{ab}}
    = \frac{1}{2}\int\limits_{V_{\mathbb{R}^d}} {{d^{d
    }}{x_1}\int\limits_{V_{\mathbb{R}^d}} {{d^{d }}{x_2}\int\limits_{ - \infty
    }^{ 0 } {d{\tau _1}\int\limits_0 ^\infty  {d{\tau _2}\left(\left\langle
    {{\cal O}_a({\tau _1},{x_1}){\cal O}_b({\tau _2},{x_2})}
    \right\rangle-\braket{{\cal O}_a({\tau _1},{x_1})}\braket{{\cal O}_b({\tau
    _2},{x_2})}\right) } } } } .  \end{equation}
	
   In order to arrive at this result one has to refer to the path-integral
  formalism, where the  overlap $\langle \psi_1|\psi_0\rangle$ can be writen in the following way
  	 \begin{equation}\label{eqPathIntQIM}
  		\braket{\psi_1|\psi_0}=\int\limits_{\varphi(\tau=0)=\tilde\varphi}\mathcal D\varphi \braket{\psi_1| \varphi}\braket{
  			\varphi|\psi_0}=\frac{1}{\sqrt{Z_0 Z_1}} \int\mathcal{D}\varphi \,e^{-\int
  			d^d x \left[\int\limits_{-\infty}^0 d\tau\mathcal{L}_0+\int\limits_{0}^\infty
  			d\tau( \mathcal{L}_0+\delta\lambda^a\mathcal{O}_a)\right]}.  
  		\end{equation}
  	Here $\ket{\varphi}$ is a generic state, while $|\tilde\varphi\rangle=|\varphi(\tau=0)\rangle$ is 
  	inserted at
  	$\tau=0$. The function $Z_0$ is the partition functions of the original theory, 
  	\begin{equation}\label{key} Z_0=\int\mathcal{D}\varphi\,
  		e^{-\int\limits_{-\infty}^\infty d\tau\int d^d x \mathcal{L}_0},
  	\end{equation}
  	while $Z_1$ represents the partition function of the deformed theory for $\tau\in(-\infty,\infty)$,
  	  \begin{equation} Z_1=\int\mathcal{D}\varphi
  		\,e^{-\int\limits_{-\infty}^\infty d\tau\int d^d x(
  			\mathcal{L}_0+\delta\lambda^a\mathcal{O}_a)}. \end{equation}
  	Now one can take the squared overlap $|\langle \psi_1|\psi_0\rangle|^2$ and expand it in powers of the couplings $\delta\lambda^a$ to obtain Eq. (\ref{eqQIMCFTsidea}), as shown in details in Appendix \ref{appB1}.
	
When considering quantum field theories in general, the overlap
(\ref{eqPathIntQIM}) could be ill defined. Since the Lagrangian, governing the
Euclidean propagation, changes discontinuously at $\tau=0$, an UV divergence
appears. As suggested in \cite{Trivella:2016brw}, one should renormalize the
theory replacing for example $\ket{\psi_1}$ by
\begin{equation}\label{key} \ket{\psi_1(\epsilon)}=\frac{e^{-\epsilon H_0}
\ket{\psi_1}}{\left(\braket{\psi_1| e^{-2\epsilon H_0}|\psi_1}\right)^{1/2}}.
\end{equation}
Here $H_0$ is the Euclidean Hamiltonian of the unperturbed theory and Eq.
(\ref{eqPathIntQIM}) now becomes
  \begin{equation}\label{eqPathIntQIM2}
    \braket{\psi_1(\epsilon)|\psi_0}=\frac{\left\langle\exp\left({-\int\limits_{\epsilon}^\infty
    d\tau\int \limits_{V_{\mathbb{R}^d}} d^d x \delta\lambda^a
    \mathcal{O}_a}\right)\right\rangle}{\left\langle
\exp\left({-\left(\int\limits_{-\infty}^{-\epsilon}+\int\limits_{\epsilon}^\infty\right)
d\tau\int \limits_{V_{\mathbb{R}^d}} d^d x \delta\lambda^a
\mathcal{O}_a}\right)\right\rangle^{1/2}}.  \end{equation}
  The expectation values are taken with respect to the unperturbed state
  $\ket{\psi_0}$. 	Consequently, equation (\ref{eqQIMCFTsidea}) changes
 upon
  expansion of (\ref{eqPathIntQIM2}) in powers of $\delta\lambda$ by the cutoff $\epsilon$ replacing the infinities along\footnote{The presence of the cutoff $\epsilon$ around $\tau=0$ is
  	necessary to address any  ultraviolet divergences in case there is
  	a discontinuity, when passing from the original to the deformed Lagrangian.} $\tau$. In order to be consistent with the original derivation of Eq. (\ref{eqQIMCFTsidea})
in the context of gauge/gravity correspondence, the
author of \cite{Trivella:2016brw} considered CFT operators of non-zero
dimension. In this case, their vacuum expectation value is
$\braket{\mathcal{O}_a}=0$. Therefore, the formula for the quantum information
metric (\ref{eqQIMCFTsidea}) reduces to
    \begin{equation}\label{eqQIMCFTsideaA} {G_{ab}}
      = \frac{1}{2}\int\limits_{V_{\mathbb{R}^d}} {{d^{d
      }}{x_1}\int\limits_{V_{\mathbb{R}^d}} {{d^{d }}{x_2}\int\limits_{
- \infty }^{ - \epsilon } {d{\tau _1}\int\limits_\epsilon ^\infty  {d{\tau
_2}\left\langle {{\cal O}_a({\tau _1},{x_1}){\cal O}_b({\tau _2},{x_2})}
\right\rangle } } } }.  \end{equation}

    Finally, one can define the quantum Fisher information metric by the real
    part of QIM, $F_{ab}=4\Re{(G_{ab})}$, while its imaginary part corresponds
    to the Berry curvature $B_{ab}=\Im{(G_{ab})}$.
		
    The derived formula (\ref{eqQIMCFTsideaA}) for QIM on the CFT side of the
    correspondence will be the relevant expression for our study of
    non-relativistic Schr\"odinger holography throughout this paper.
	
  \subsection{Bulk holographic construction} Early calculations of the Fisher
  information metric on the gravity side with exactly marginal operators of the
  deformation  have been conducted in \cite{MIyaji:2015mia,Bak:2015jxd}. Bulk
  holographic study for AdS$_{d+1}$ space lead to the development of
  a specific perturbative method \cite{Trivella:2016brw}, which allows us to
  deal with any primary operator with conformal dimension $\Delta$, provided
  that
  \begin{equation}\label{eqSomeCond} 2\Delta>d+1.  \end{equation}
	
  The basic idea of the suggested technique is to reinterpret the right hand
  side of equation (\ref{eqPathIntQIM}) as a combination of partition
  functions, namely
  \begin{equation} \langle \psi_1|\psi_0\rangle=\frac{Z_2}{\sqrt{Z_0 Z_1}},
  \end{equation}
  were $Z_2$ is the partition function for the deformed theory only for
  $\tau>0$. The latter has been explicitly defined in the numerator of
  \eqref{eqPathIntQIM}. Now, one can consider the large $N$ limit on the
  gravity side, where $Z_k=e^{-I_k}$, $k=0,1,2$, with $I_k$ being the on-shell
  action of the gravity solution dual to the corresponding field theory
  configuration. Therefore, one can write
  \begin{equation} \langle {\psi _1}|{\psi _0}\rangle  = \frac{{{Z_2}}}{{\sqrt
  {{Z_0}{Z_1}} }} = {e^{\frac{1}{2}({I_0} + {I_1}) - {I_2}}}.  \end{equation}
  The deformation of the CFT by a single primary $\mathcal{O}$ induces an
  interaction term  $\delta\lambda \mathcal{O}$ with a coupling $\delta\lambda$
  in the CFT Lagrangian, thus changing the initial dual bulk gravitational
  action $I_0$ by $I_k=I_0+\delta I_k(\delta\lambda)$ with $\delta I_0=0$.
  Hence, one has to compute
  \begin{equation}\label{eqAlaBala} \langle {\psi _1}|{\psi _0}\rangle
    = \frac{{{Z_2}}}{{\sqrt {{Z_0}{Z_1}} }} = {e^{\frac{1}{2}({I_0} + {I_1})
    - {I_2}}} = {e^{\frac{1}{2}({I_0} +\delta I_0 + {I_0} + \delta {I_1})
- {I_0} - \delta {I_2}}} = {e^{\frac{1}{2}\delta {I_1} - \delta {I_2}}}.
\end{equation}

  As a result,  the final computation reduces to finding the variations $\delta
  I_1$ and $\delta I_2$ of the on-shell gravitational action. Let us assume
  that the bulk spacetime dynamics is governed by the action
  \begin{equation}\label{eqtheAction} I =  - \frac{1}{{{\kappa
    ^2}}}\int\limits_{\cal M}{d^{D}}x\sqrt {|g|} \left(
  {\frac{1}{2}(R-2\Lambda) - \frac{1}{2}{g^{\mu \nu }}{\partial _\mu }\phi
{\partial _\nu }\phi  - \frac{1}{2}{m^2}{\phi ^2} + V(\phi)} \right)
+ {I_{\partial {\cal M}}}, \end{equation}
  where $\phi(x)$ is a massive scalar field probing the background geometry
  $g_{\mu \nu}(x)$.  The field $\phi(x)$ attains different profiles for
  $k=0,1,2$. In particular, for the computation of $Z_0$, we notice that the
  massive field is turned off ($\phi_0=0$), thus the dual solution is the
  initial background geometry with metric $g^{(0)}_{\mu\nu}$, i.e.
  $Z_0=\exp\left({-I_0[\phi_0,g^{(0)}_{\mu\nu}}]\right)$. On the other hand,
  the profiles  $\phi_{1,2}(x)$ of the scalar field for $I_1$ and $I_2$ will in
  general be spacetime dependent and can be calculated by the corresponding
  bulk-to-boundary propagator. This is shown explicitly in the case of
  Schr\"odinger spacetime in Section \ref{sec4} and Appendix \ref{appB}. Since
  we are interested in perturbative solutions in lower powers of
  $\delta\lambda$,  we employ the following transformations of the fields
  \begin{equation} \phi (x) = {\phi _0}(x) + \varphi (x) \delta\lambda ,\qquad
  {g_{\mu \nu }}(x) = g_{\mu \nu }^{(0)}(x) + h_{\mu \nu }^{}(x)
{\delta\lambda^2}.  \end{equation}
  Here $\varphi(x)$ and $h_{\mu \nu }^{}(x)$ are the corrections to the scalar
  field and the bulk metric acquired after turning on the deformation in the
  dual CFT.  Notice that the metric receives corrections at order
  $ {\delta\lambda ^2}$ since the scalar field enters quadratically in
  the  Einstein
  field equations. Hence, the variation of the bulk action $\delta I$ now can
  be computed in powers of $\delta\lambda$, 
  \begin{align}\label{eqVariationoftheAction} \nonumber \delta I &= I[\phi
    ,{g_{\mu \nu }}] - {I_0}[{\phi _0},g_{\mu \nu }^{(0)}] = I[{\phi _0}(x)
    + \varphi (x) \delta\lambda ,g_{\mu \nu }^0(x) + h_{\mu \nu }^{}(x)
    {\delta\lambda ^2}] - {I_0}[{\phi _0},g_{\mu \nu }^{(0)}]\\\nonumber &=
    \delta\lambda\int {\frac{{\delta I}}{{\delta \phi (x)}}\left|
    \begin{array}{l} \\ _{ \delta\lambda  = 0} \end{array} \right.} \varphi (x)
    + \delta\lambda ^2\int {\frac{{{\delta ^2}I}}{{\delta \phi (x)\delta \phi
      (y)}}\left| \begin{array}{l} \\ _{ \delta\lambda  = 0} \end{array}
      \right.} \varphi (x)\varphi (y) \\ &+ \delta\lambda ^2\int {\frac{{\delta
I}}{{\delta {g_{\mu \nu }}(x)}}\left| \begin{array}{l} \\ _{\delta\lambda  = 0}
\end{array} \right.} h_{\mu \nu }^{}(x) +  \mathcal{O}(\delta\lambda^3) \approx
\delta\lambda ^2\int {\frac{{{\delta ^2}I}}{{\delta \phi (x)\delta \phi
(y)}}\left| \begin{array}{l} \\ _{ \delta\lambda  = 0} \end{array} \right.}
\varphi (x)\varphi (y).  \end{align}
  The first and the last term vanish due to the field equations of motion.
  Higher order contributions are not taken into account, due to the fact that
  we are working in the probe limit and backreaction on the background geometry
  is considered negligible. Hence, we are in a situation where the scalar field
  probes the unperturbed background. Notice also that the boundary term of
  equation (\ref{eqtheAction}) cancels by the boundary terms coming from the
  integration by parts in obtaining the first and the third terms of Eq.
  (\ref{eqVariationoftheAction}). Therefore, we can write 
  \begin{equation}\label{eqdI} \delta I_k = \frac{1}{{2{\kappa ^2}}}\! \int\!
    d^{d + 1}x\, \sqrt {|g|} \left( g^{(0)\mu \nu } \partial_\mu \phi_k\,
    \partial _\nu \phi_k + {m^2}\phi _k^2\right) =\frac{1}{{2{\kappa ^2}}}\!
    \int\! {d^d}x\, \sqrt {|\gamma |}\, n_\mu\, g^{(0)\mu \nu } \phi_k\,
    {\partial _\nu }{\phi _k}, \end{equation}
  with $n_\mu$ being the unit normal vector and $\gamma$ being  the determinant of the
  induced metric on the boundary. Furthermore, $\phi_k(x)$ are the scalar field
  configurations dual to the operators of the corresponding deformed and
  undeformed CFTs, while probing the fixed background $g^{(0)}_{\mu\nu}$.
  Clearly, one can obtain these profiles by using the boundary-to-bulk
  propagator, which we will show in the following sections. The
  next step is
  to write the overlap \eqref{eqAlaBala} as
  \begin{equation}\label{eqAla} \langle {\psi _1}|{\psi _0}\rangle
  = \frac{{{Z_2}}}{{\sqrt {{Z_0}{Z_1}} }} = \exp \left( {\frac{1}{2}\delta
{I_1} - \delta {I_2}} \right).  \end{equation}
	
  Finally, after expanding the exponent in \eqref{eqAla} up to first order in
  $\delta I_{1,2}$ and comparing to \eqref{eqBala}, one finds an expression for
  the holographic information metric
  \begin{equation}\label{eqHolFIM} {G_{\lambda \lambda }} = -\,\frac{1}{{\delta
  {\lambda ^2}}}\left( \frac{{\delta {I_1}}}{2}-\delta {I_2} \right).
\end{equation}
 The real part of this expression leads to the holographic Fisher
  information metric (HFIM)
  \begin{equation}\label{key} F_{\lambda\lambda}=4\Re(G_{\lambda\lambda}).
  \end{equation}

  In what follows, we are going to calculate the quantum Fisher information
  metric (QFIM) and its dual holographic counterpart HFIM for a holographic
  system in Schr\"odinger spacetime, according to the computational procedures
  presented in this section.
	
  \section{Dual CFT quantum Fisher information metric}\label{sec3}
	
 As is evident from Eq. \eqref{eqQIMCFTsideaA}, in order to find the QFIM, we
 must compute the two-point correlation function of operators with dimensions
 $\Delta$ from the dual conformal field theory. In many cases, this can be
 achieved within the framework of the gauge/gravity correspondence. One has to
 look at the dynamics of a massive scalar field (\ref{eqtheAction}) propagating
 in the bulk geometry from the gravity side. 
		
  In our case, we consider a strongly coupled $(d+1)$-dimensional conformal
  field theory with non-relativistic invariance\footnote{ It is conjectured to
  be a specific non-local field theory, namely dipole field theory (see
Appendix \ref{app0} for more information).}. It is assumed to be dual to
a ($d+3$)-dimensional Schr\"odinger spacetime with Lorentzian line element
given by\cite{Son:2008ye, Balasubramanian:2008dm}
  \begin{equation}\label{eqSchrMetric} ds^2_{Schr_{d+3}} = {L^2}\left( { -\,
  \frac{{d{t^2}}}{{{r^4}}} + \frac{{2d\xi dt + d{{\vec x}^{\,2}}}}{{{r^2}}}
  + \frac{{d{r^2}}}{{{r^2}}}} \right).  \end{equation}
  The boundary of the background (\ref{eqSchrMetric}) is at $r = 0$ and the
  generator associated with translations along the compact $\xi$ direction can
  be identified with the mass operator $M = i\partial_\xi$. The latter is not
  a geometric dimension in the usual sense. Each  operator of the boundary
  theory can have a fixed momentum (‘particle number’) conjugate to $\xi$,
  where the compactification of $\xi$  is usually taken so that the spectrum of
  possible momenta is discrete. Hence, the effective dimension of the boundary
  CFT has to be $(d + 1)$, which is coordinazied by $(t, \vec x)$ or
  ($\tau,\vec x$) in the Euclidean case. Now, the 2-point function between
  primary operators from the boundary gauge theory with conformal dimension
  $\Delta$ can be computed by a standard holographic procedure implemented in
  \cite{Volovich:2009yh, Leigh:2009eb}. The explicit form of the correlator is
  \begin{equation} {A_{12}} = \left\langle {\mathcal{O}({\tau_2},{{\vec
    x}_2})\mathcal{O}({\tau_1},{{\vec x}_1})} \right\rangle  = \Delta\,
    c_\Delta\, \theta ({\tau_2} - {\tau_1}){\left( {\frac{1}{{{\tau_2}
    - {\tau_1}}}} \right)^\Delta } {\exp\left({i\frac{M (1+i
    \varepsilon)}{2}\frac{{{{({{\vec x}_2} - {{\vec x}_1})}^2}}}{{{\tau_2}
    - {\tau_1}}}}\right)}, \end{equation}
  where $c_\Delta$ is a normalization constant,
  \begin{equation}\label{eqNormC} {c_\Delta } = \frac{{i{{\left( {\frac{M}{2}}
  \right)}^{\Delta  - 1}}{e^{ - i\pi \frac{\Delta }{2}}} }}{{{\pi ^{d/2}}\,
  \Gamma \left( {\Delta  - \frac{d}{2} - 1} \right)}}, \end{equation}
  and $\Delta$ the conformal dimension, given by
  \begin{equation} \Delta
  = 1+\frac{d}{2}+\sqrt{\left(1+\frac{d}{2}\right)^2+m^2+M^2}.  \end{equation}
  Here, $d$ is the dimension of $\vec x$ space, $\theta(\tau_2-\tau_1)$ is the
  unit step function for an Euclidean time interval and $M$ is a quantized
  momentum along the compact direction $\xi$ with radius $1/M$. We will
  consider the case $\tau_1\leq\tau_2$, thus QIM for a single marginal
  deformation should be given by

  \begin{equation}\label{eqQIMCFTside} G_{\lambda \lambda }^{(CFT)}
    = \frac{1}{2} \int\! {d^{d} x_1 \int\! {d^{d} x_2 \int\limits_{-\infty}^{-
    \epsilon}\! {d\tau _1 \int\limits_\epsilon^\infty\!  {d\tau _2 \left\langle
    {{\cal O}({\tau _1},{x_1})\, {\cal O}({\tau _2},{x_2})} \right\rangle
    } } } }, \end{equation}
  where we have also taken into account that $\braket{\mathcal{O}}=0$ for an
  operator of non-zero dimension. Explicitly, one has
  \begin{equation} \label{eqQFIMGauss} G_{\lambda \lambda }^{(CFT)}
    = \frac{\Delta\, c_\Delta}{2} \int\! {{d^{d }}{x_1} \int\! {{d^{d}}{x_2}
    \int\limits_{-\infty}^{- \epsilon}\! {d{\tau_1}
\int\limits_\epsilon^\infty\!  {d\tau_2\, \frac{{e^{i\frac{M(1+i
\varepsilon)}{2}\frac{{{{({{\vec x}_2} - {{\vec x}_1})}^2}}}{{{\tau_2}
- {\tau_1}}}}}}{(\tau_2 - \tau_1)^{\Delta}} } } } }, \end{equation}
  where $\epsilon$ is a regulator near $\tau=0$, which is different from the
  regulator $\varepsilon$ in the correlation function.  The integral over $\vec
  x$ space is Gaussian and can be easily computed  by Eq. (\ref{eqGaussianInt})
  \begin{equation} {I_x} = \int\! {{d^{d }}{x_1} \int\! {{d^{d}}{x_2}\,
    {e^{i\frac{M(1+i \varepsilon)}{2}\frac{{{{({{\vec x}_2} - {{\vec
x}_1})}^2}}}{{{\tau_2} - {\tau_1}}}}}} }  = {e^{\frac{{i\pi }}{2}(1
- d/2)}}{(2\pi )^{\frac{{d}}{2}}}{M^{\frac{{- d}}{2}}(1+i
\varepsilon)^{-\frac{d}{2}}} (\tau_2-\tau_1)^\frac{d}{2}{V_{{\mathbb{R}^{d}}}},
\end{equation}
  where $V_{\mathbb{R}^{d}}$ is the volume of $\mathbb{R}^{d}$ space.  The
  integrals over $\tau_1$ and $\tau_2$ now take the form
  \begin{equation}\label{eqTimeInt} I_t = \int\limits_{-\infty }^{
    - \epsilon}\! {d{\tau_1} \int\limits_\epsilon^\infty\!  {d\tau_2\, (\tau_2
  - \tau_1)^{\frac{d}{2} - \Delta }} }  = \frac{{{2^{\frac{d}{2} - \Delta
  + 4}}}}{{(d - 2\Delta  + 2)(d - 2\Delta  + 4)}}\, \epsilon ^{2+ \frac{d}{2}
- \Delta }, \end{equation}
  with convergence condition given by
  \begin{equation}\label{eqDivCondQuantumFisher} 2\Delta>d+4, \end{equation}
  which falls within the range specified in Eq. (\ref{eqSomeCond}).  Therefore,
  the QIM in the dual to the Schr\"odinger spacetime dipole CFT is given by
  \begin{equation} {G_{\lambda \lambda }^{(CFT)}} =\frac{{\Delta {2^{d
    - 2\Delta  + 3}}{e^{- \frac{{i\pi }}{4}(d + 2\Delta )}}{M^{ \Delta
    - \frac{d}{2}  - 1}}{V_{{\mathbb{R}^d}}}}}{{(d - 2\Delta + 4) \,\Gamma
\left( {\Delta  - \frac{d}{2}} \right) {(1+i
\varepsilon)^{\frac{d}{2}}}}}{\,\epsilon ^{2 + \frac{d}{2} - \Delta }}.
\end{equation}
  We can now safely turn off the regulator $\varepsilon=0$, which is equivalent
  to switching back to real time $t$ \cite{Volovich:2009yh}, hence
  \begin{equation}\label{eqCFTDualFIM} {G_{\lambda \lambda }^{(CFT)}} =C
  \epsilon^{1-a}, \end{equation}
  where $C$ is a constant normalization factor given by
  \begin{equation}\label{eqCFTDualFIMC} C=\frac{{\Delta {2^{d - 2\Delta
    + 3}}{e^{- \frac{{i\pi }}{4}(d +2\Delta )}}{M^{ \Delta - \frac{d}{2}
    - 1}}{V_{{\mathbb{R}^d}}}}}{{( d - 2\Delta + 4) \, \Gamma \left( {\Delta
    - \frac{d}{2}} \right)}}.  \end{equation}
  Here, we have also defined the parameter
  \begin{equation}\label{eqA} a=\Delta-\frac{d}{2}-1, \end{equation}
  to better outline the divergence structure of the QIM.  Since the
    divergence parameter $\epsilon$ is a real parameter, the only difference
    between QIM and QFIM resides only in their normalization constants, 
\begin{equation}\label{eqQFIM}
F_{\lambda\lambda}=4\Re(G_{\lambda\lambda})=4\Re(C)
\epsilon^{1-a}=F\epsilon^{1-a}.  \end{equation}

We can now proceed with the computation of the holographic Fisher information
metric in the dual gravitational theory.
	
\section{Bulk holographic Fisher information metric}\label{sec4}
	
We focus on the dynamics of the massive scalar field $\phi(x)$ on the gravity
side. In our setup we have the following picture. The initial theory is a CFT
$\mathcal{L}_0$ with primary operators dual to a probe scalar field $\phi_0(x)$
in the bulk defined for $\tau\in (-\infty,+\infty)$\footnote{Note that this
picture is a bit different from the one in Section (\ref{sec2-1}). When
considering the dual gravitational theory we have to consider the full range of
the CFT theory $\mathcal{L}_0$ for $\tau\in(-\infty, +\infty)$. The same is
valid also for the deformed theory with $\mathcal{L}_1$, which now should also
be considered for $\tau\in(-\infty, +\infty)$. Finally, a new theory with
$\mathcal{L}_2=\mathcal{L}_1$ only for $\tau\in(0, +\infty)$ is required for
consistency of the perturbative bulk holographic method.}. On the other hand,
for the same $\tau\in (-\infty,+\infty)$, there is another CFT with Euclidean
Lagrangian $\mathcal{L}_1=\mathcal{L}_0+\delta \lambda \mathcal{O}$, which is
a deformation of the original theory by $\delta \lambda \mathcal{O}$. Its
operator content is dual to a new probe scalar field $\phi_1(x)$ in the bulk.
Finally, we consider a third theory with $\mathcal{L}_2=\mathcal{L}_1$, defined
only for $\tau>0$. It is produced  by deforming the initial CFT at an Euclidean
time $\tau=0$ and its content is dual to a bulk scalar field $\phi_2(x)$. 
		
As a consequence of Eq. \eqref{eqHolFIM} the computation of the holographic
Fisher information metric requires the profiles only for $\phi_1(x)$ and
$\phi_2(x)$. These solutions can be easily found by the method of Green's
functions (bulk-to-boundary propagators).  In this case, the bulk-to-boundary
propagator $K(r,\vec x,\tau;{{\vec x}_1},{\tau_1})$ in Schr\"odinger spacetime
is given by \cite{Volovich:2009yh} 
  \begin{equation}\label{key} K(r,\vec x,\tau;{{\vec x}_1},{\tau_1})
    =c_\Delta\, \theta (\tau - \tau_1)\frac{{{r^\Delta }}}{{{{(\tau
    - {\tau_1})}^\Delta }}}\, {\exp\! \left({\frac{i M}{2}\frac{{{r^2}
    + {{(\vec x - {{\vec x}_1})}^2}}}{{\tau - {\tau_1}}}}\right)},
  \end{equation}
  where the normalization constant $c_\Delta$ has been defined in Eq.
  (\ref{eqNormC}). We choose to perform the integration over $x_1$ and the
  Euclidean time $\tau_1$  for $\tau_1\leq\tau$. The detailed computation has
  been presented in Appendix \ref{appB}. Hence, one can write the solution for
  $\phi_1(x)$ in the deformed theory as\footnote{The source function $\hat \phi
  _0(\vec x_1,\tau_1)=1$, which can be seen from the asymptotic expansion of
the Bessel functions, defining the propagator near the boundary at $r=0$, see
\cite{Balasubramanian:2008dm}. }
  \begin{align}\label{eqPhi1} {\phi _1}(r,\vec x,t) &= \delta\lambda \int\! d^d
    x_1 \int\limits_{-\infty }^\tau\! d{\tau_1} \, K(r,\vec x,\tau;{{\vec
    x}_1},{\tau_1})\, \hat \phi _0(\vec x_1,\tau_1)= i\delta\lambda\, e^{-
  \frac{{i\pi d}}{2}} r^{d - \Delta  + 2}.  \end{align}
It is valid only if the following convergence conditions for the integrals hold
  \begin{equation}\label{eqConvCond1} a>0,\quad M>0, \end{equation}
  where $a$ is given by Eq.\eqref{eqA} and $M$ is the non-relativistic momentum
  associated to the compact Killing direction $\xi$.  The convergence condition
  $a>0$ is weaker than (\ref{eqDivCondQuantumFisher}), which translates to
  $a>1$. However matching HFIM to QFIM, as shown in Section \ref{sec5},
  requires $a>1$ in the bulk as well.  One also notes that ${\phi _1}(r,\vec
  x,\tau) \equiv {\phi _1}(r)$ now becomes a function only of $r$, which
  considerably simplifies the subsequent computation of the HFIM. Furthermore,
  the choice of $\tau\to+\infty$ as an upper limit of the integral does not
  change the final result in (\ref{eqPhi1}). 
	
  The field $\phi_2$, for the deformed theory at $\tau>0$, is given by
  \begin{align}\label{key} {\phi _2}(r,\vec x,t) =\delta\lambda \int\! d^d x_1
    \int\limits_0^\tau\! d\tau_1\, K(r,\vec x,\tau;{{\vec x}_1},{\tau_1})\,
    \hat\phi _0(\vec x_1,\tau_1)=\frac{i \delta\lambda\, e^{-\frac{i \pi
    d}{2}}}{\Gamma(a)}\, r^{d-\Delta+2}\, \Gamma\left(a,\frac{\mu}{t}\right),
  \end{align}
  where in the final step we have restored the real time $\tau\to -i t$, so
  that subsequent calculations with the incomplete gamma functions will attain
  the correct properties. The convergence condition for $\phi_2$ is $a>0$ and
  we have also introduced the real parameter
  \begin{equation}\label{key} \mu  = \frac{M r^2}{2}.  \end{equation}
  This parameter is a convenient choice for a regulator along the holographic
  coordinate $r$, which we will employ in the sections below. 
	
  With the profiles of the fields $\phi_{1,2}$ at hand we can proceed with the
  computation of the HFIM integrals according to (\ref{eqdI}), adopted for the
  Schr\"odinger spacetime (\ref{eqSchrMetric}),
  \begin{equation}\label{key} \delta I_k = \frac{1}{{2\kappa }}\mathop {\lim
    }\limits_{r \to \tilde \varepsilon } \int\! {{d^d}x\left(\, {\int\limits_{
  - {T}}^{-\epsilon}\! {dt\sqrt {|\gamma |}\, n_\mu\, {g^{\mu \nu }} {\phi
  _k}\, {\partial _\nu }{\phi _k} + \int\limits_\epsilon^{T}\!   dt\sqrt
  {|\gamma |}\, n_\mu\, {g^{\mu \nu }}{\phi _k}\, {\partial _\nu }{\phi _k}
  } } \right)} .  \end{equation}
  Here $\gamma$ is the metric on the boundary, $\tilde \varepsilon>0$ is the
  regulator in the holographic direction $r$ near the boundary $r\to 0$,
  $n_\mu$ is the normal outward vector to the boundary, $\epsilon>0$ is
  a regulator around $t=0$, and ${T} $ is a cut-off at temporal infinity. On
  the boundary only the component $n_r$ is non-zero, thus  $\sqrt{|\gamma|}\,
  n_r g^{rr}=L^{d}\, r^{-1-d}$ (see Appendix \ref{appB2}). However, one has to
  account for the fact that $\phi_1$ and $\phi_2$ span different ranges along
  $t$, which leads to
  \begin{align}\label{eqDI1} &\delta {I_1}
    = \frac{{{L^{d}}{V_{{\mathbb{R}^d}}}}}{{2\kappa }}\mathop {\lim
    } \limits_{r \to \tilde \varepsilon } \left(\,
  \int\limits_{-{T}}^{-\epsilon}\! dt \, r^{-1-d}\, \phi_1(r)\, {\partial _r
}{\phi _1(r)} + \int\limits_\epsilon^{T}\!  dt\, r^{-1-d}\, \phi_1(r)\,
{\partial _r }{\phi _1(r)}  \right) , \\\label{eqDI2} &\delta {I_2}
= \frac{{{L^{d}}{V_{{\mathbb{R}^d}}}}}{{2\kappa }}\mathop {\lim }\limits_{r \to
\tilde \varepsilon } {{ \int\limits_\epsilon ^{T}\!   dt\, r^{-1-d}\,
\phi_2(t,r)\, {\partial _r }{\phi _2(t,r)} } } .  \end{align}
  Now, Eq. (\ref{eqHolFIM}) for the holographic Fisher information metric leads
  to the expression 
  \begin{equation}\label{key} {G_{\lambda \lambda }^{Bulk}} = -\,
    \frac{1}{{\delta {\lambda ^2}}}\left( \frac{{\delta {I_1}}}{2}-\delta {I_2}
    \right) = \frac{{{L^{d}}{V_{{\mathbb{R}^d}}}}}{{2\kappa }}\mathop {\lim
  }\limits_{r \to \tilde \varepsilon } \left( {{c_0}{J_0}\, r^{d-2 \Delta +2}
+ {c_1}{J_1} + {c_2}{J_2}\, r^{d-2 \Delta +2}} \right), \end{equation}
  with the following integrals
  \begin{align}\label{key} &{J_0} = {\int\limits_{-T} ^{-\epsilon}\! dt
    + \int\limits_{\epsilon}^{T}\!  {dt}  } = 2(T - \epsilon),\\ &{J_1}
    =  \int\limits_{\epsilon}^{T}\!   {dt}\, t^{-a}\, e^{-\frac{\mu }{{t}}}\,
    \Gamma\! \left( {a,\frac{\mu }{{t}}} \right)
    = \int\limits_{1/T}^{1/\epsilon}\! dx\, x^{a-2}\, e^{-\mu x}\, \Gamma(a,\mu
    x),\\ &{J_2} =  \int\limits_{\epsilon}^{T}\! {dt} \,\, {\Gamma ^2}\! \left(
  {a,\frac{\mu }{{t}}} \right), \end{align}
  and the corresponding coefficients
  \begin{align}\label{key} {c_0} = \frac{e^{-i \pi  d}}{2} (d-\Delta+2) , \quad
  {c_1} = \frac{M^a\, e^{-{id \pi }}}{2^{a-1}\, \Gamma^2(a) }  , \quad{c_2}
= -\, \frac{2c_0}{\Gamma^2(a)}.  \end{align}
  The non-trivial solutions to these integrals are presented in Appendix
  \ref{appB}. The final expression for the HFIM yields\footnote{ In
  this section we prefer to directly call ${G_{\lambda \lambda }^{Bulk}}$ HFIM,
which in the correct asymptotic limits will be shown to reduce to QIM and thus
equivalently to QFIM.}
  \begin{align}\label{eqFisherBulk} \nonumber {G_{\lambda \lambda }^{Bulk}} &=
    \frac{a_0}{\mu^{a}} (T - \epsilon ) \,+\, a_1 \frac{T^{2 - a} e^{
    -\frac{\mu}{T}}}{\mu} \,  \Gamma\! \left( a-2,\frac{\mu}{T} \right) \,+\,
    \frac{a_2}{\mu^{a-1}}\, \Gamma^2\! \left( a-2,\frac{\mu}{T}
    \right)\\\nonumber &+ \frac{a_3}{\mu^{a-1}}\, \Gamma\! \left( 2a
  - 4,\frac{2\mu}{T} \right) \,+\, \frac{a_4}{\mu^{a-1}}\, \Gamma\! \left(
2a-2,\frac{2\mu}{T} \right) \,+\, \frac{a_5}{\mu^{a-1}}\, \Gamma\! \left(
2a-3,\frac{2\mu}{T} \right)\\\nonumber & + a_6 \frac{T}{\mu^a}\, \Gamma^2\!
\left( a,\frac{\mu }{T} \right) \,+\, \frac{a_7}{\mu^{a-1}}\, \Gamma^2\! \left(
a-1,\frac{\mu }{T} \right)\\\nonumber &+ b_1 \frac{\epsilon^{2-a}}{\mu}\, e^{
-\frac{\mu}{\epsilon}}\, \Gamma\! \left( a-2,\frac{\mu}{\epsilon} \right) \,+\,
\frac{b_2}{\mu^{a-1}}\, \Gamma^2\! \left( a-2,\frac{\mu }{\epsilon}
\right)\\\nonumber & + \frac{b_3}{\mu^{a-1}}\, \Gamma\! \left( 2a-4,\frac{2\mu
}{\epsilon} \right) \,+\, \frac{b_4}{\mu^{a-1}}\, \Gamma\! \left(
2a-2,\frac{2\mu}{\epsilon } \right) \,+\, \frac{b_5}{\mu^{a-1}}\, \Gamma\!
\left( 2a-3,\frac{2\mu} {\epsilon} \right)\\ & + b_6 \frac{\epsilon}
{\mu^{a}}\, \Gamma^2\! \left( a,\frac{\mu}{\epsilon} \right) \,+\,
\frac{b_7}{\mu^{a-1}}\, \Gamma^2\! \left( a-1,\frac{\mu}{\epsilon} \right).
\end{align}
	
  This result is equipped with three different divergences near the
  Schr\"odinger boundary at $r=0$, namely $\epsilon\to 0$, $T\to\infty$ and
  $\mu\sim \tilde \varepsilon^2\to 0$. In the bulk spacetime the parameter
  $\mu$ is finite, except at $\mu\to \infty$, as schematically depicted on Fig.
  \ref{fig1}.  Furthermore, one notes the relation $a_i=-b_i$ between the
  coefficients $a_i$ and $b_i$, $i=1,\dots, 7$, where
	
  \begin{align} \nonumber &{a_0} = \frac{{{L^{d
    }}{V_{{\mathbb{R}^d}}}}}{2^{a+1}\kappa } M^a \left[ {{2c_0} + {c_2}\left(
      {{\Gamma ^2}\left( a \right) - 2\Gamma (2a)\, \mathcal{B}_{1/2} (a,a)}
      \right)} \right],\\\nonumber &{a_1} =  - {b_1} = \frac{{{L^{d
}}{V_{{\mathbb{R}^d}}}}}{{2\kappa }}{c_1}(a - 1)(a - 2),\\\nonumber &{a_2}
=  - {b_2} = \frac{{{L^{d }}{V_{{\mathbb{R}^d}}}}}{{4\kappa }}{c_1}{(a - 1)}(a
- 2)^2,\\\nonumber &{a_3} =  - {b_3} =  - \frac{{{L^{d
}}{V_{{\mathbb{R}^d}}}}}{{2^{2a-3}\kappa }} c_1(a - 1)(a - 2),\\\nonumber
&{a_4} =  - {b_4} = \frac{{{L^{d }}{V_{{\mathbb{R}^d}}}}}{{2^{3a-1}\kappa }}
         \left(2^a {c_1}-2 {c_2} M^a\right),\\\nonumber &{a_5} =  - {b_5}
         = \frac{{{L^{d }}{V_{{\mathbb{R}^d}}}}}{{2^{2a-2}\kappa }} c_1 (a
         - 1),\\\nonumber &{a_6} =  - {b_6} = \frac{{{L^{d
       }}{V_{{\mathbb{R}^d}}}}}{{2^{a+1}\kappa }} M^{a} c_2,\\ &{a_7}
     =  - {b_7} =  - \frac{{{L^{d }}{V_{{\mathbb{R}^d}}}}}{{2^{a+1}\kappa }}
   M^{a} c_2(a - 1), \end{align}
  and $\mathcal{B}_{1/2}\left( {a,a} \right)$ is the incomplete beta function.
  We have explicitly kept the different names for the
  coefficients $a_i$ and $b_i$ to reflect the different asymptotic behaviour of the arguments of the
  incomplete gamma functions.
	
  \begin{center} \begin{figure}[H] \begin{center}
    \begin{tikzpicture}[x={(0,1cm)},y={(1 cm,0)}] \path [domain=-2:0,pattern
      color=white, pattern=north east lines, fill opacity=.5] (0,-1) --
      (2,-1)--(2,3) -- plot  (\x,-\x*\x) --cycle ;  \draw
        [red,domain=-2.5:2.5,samples=100, thick]  plot (\x,-\x*\x)
        ;  \draw[thin,blue] (3,-6.27) -- (-3,-6.27) ; 
				
        \draw[->,thick] (0,-6.27) -- (0,0);  
				
        \node at (0.0,0.75) {\footnotesize $\mu\to\infty$};   \node at (0,-6.9)
          {\footnotesize $\mu\to0$};  \node at (1,-3.9) {\footnotesize
            (d+3)-dim Bulk};   \node at (1.6,-7.3) {\footnotesize (d+1)-dim};
            \node at (1.2,-7.3) {\footnotesize Boundary}; \node at (0.8,-7.3)
            {CFT}; \end{tikzpicture} \end{center} \caption{A schematic
            depiction of the Schr\"odinger spacetime and its boundary along
          the holographic direction $r$ ($\mu\sim r^2$). We have three
        important sectors capturing different divergence structures. To the left is the $(d+1)$-dimensional boundary at
      $r=0$, where the dual CFT lives. The middle region defines the
    $(d+3)$-dimensional bulk of the considered space, where $r\neq0$ is finite,
  and to the right is the limit $r\to\infty$.}\label{fig1}
\end{figure}
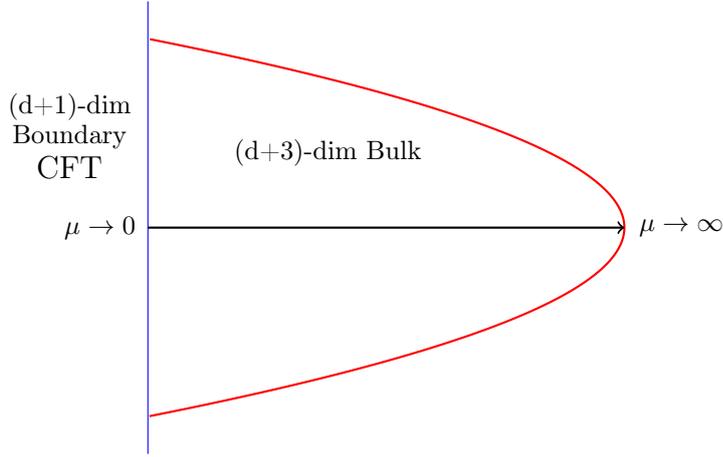 \end{center}
	
  \section{Reduction of HFIM to the dual QFIM on the boundary} \label{sec5}
	
  In order to compare HFIM from Eq. (\ref{eqFisherBulk}) to its CFT dual from
  Eq. (\ref{eqCFTDualFIM}) we have to  consider the divergence structure of
  HFIM near the boundary at $r=0$. In this case, we have effectively two
  competing divergences, namely $\mu\to 0$ when approaching $r=0$ along $r$,
  and $\epsilon\to 0$ along $t$ near $t=0$. Therefore, we can look at different
  situations, e.g. one in which $\mu$ goes to zero asymptotically faster than
  $\epsilon$, and the other case, where $\epsilon$ goes faster to zero than
  $\mu$. In both cases we have two possibilities for the cut-off $T$, i.e.
  $T\to \infty$ or finite $T$. Let us rewrite HFIM from Eq.
  (\ref{eqFisherBulk}) in the form
  \begin{align}\label{eqFisherFull} \nonumber
    \mathcal{G}_{\lambda\lambda}&={G_{\lambda \lambda }^{Bulk}} \epsilon ^{a-1}
    = {a_0} \frac{T \epsilon^{a-1}}{\mu^a} \,-\, {a_0} \frac{\epsilon
  ^a}{\mu^a}\\\nonumber &+ {a_1} \frac{T^{2-a} e^{-\frac{\mu}{T}} \epsilon
^{a-1} }{\mu}\, \Gamma\! \left( {a - 2,\frac{\mu }{T}} \right) \,+\, a_2
\frac{\epsilon^{a-1}}{\mu^{a-1}}\, \Gamma^2\! \left( {a - 2,\frac{\mu }{T}}
\right)\\\nonumber &+ a_3 \frac{\epsilon^{a-1}}{\mu^{a-1}}\, \Gamma\! \left(
{2a - 4,\frac{{2\mu }}{T}} \right) \,+\, a_4 \frac{\epsilon^{a-1}}{\mu^{a-1}}\,
\Gamma\! \left( {2a - 2,\frac{{2\mu }}{T}} \right)\\\nonumber &+ a_5
\frac{\epsilon^{a-1}}{\mu^{a-1}}\, \Gamma\! \left( {2a - 3,\frac{{2\mu }}{T}}
\right) \,+\, a_6 \frac{T \epsilon^{a-1}}{\mu^a}\,  \Gamma^2\! \left(
{a,\frac{\mu }{T}} \right) \,+\, a_7 \frac{\epsilon^{a-1}}{\mu^{a-1}}\,
\Gamma^2\! \left( {a - 1,\frac{\mu }{T}} \right)\\\nonumber &+ b_1 \frac{e^{
-\frac{\mu }{\epsilon}} \epsilon}{\mu} \, \Gamma\! \left( {a - 2,\frac{\mu
}{\epsilon }} \right) \,+\, b_2 \frac{\epsilon^{a-1}}{\mu^{a-1}}\, \Gamma^2\!
\left( {a - 2,\frac{\mu }{\epsilon }} \right)\\\nonumber &+ b_3
\frac{\epsilon^{a-1}}{\mu^{a-1}}\, \Gamma\! \left( {2a - 4,\frac{{2\mu
}}{\epsilon}} \right) \,+\, b_4 \frac{\epsilon^{a-1}}{\mu^{a-1}}\, \Gamma\!
\left( {2a - 2,\frac{{2\mu }}{\epsilon }} \right)\\ &+ b_5
\frac{\epsilon^{a-1}}{\mu^{a-1}}\, \Gamma\! \left( {2a - 3,\frac{{2\mu
}}{\epsilon}} \right) \,+\, b_6 \frac{\epsilon^a}{\mu^a}\, \Gamma^2\! \left(
{a,\frac{\mu }{\epsilon }} \right) \,+\, b_7 \frac{\epsilon^{a-1}}{\mu^{a-1}}\,
\Gamma^2\! \left( {a - 1,\frac{\mu }{\epsilon }} \right).  \end{align}
  Our goal is to compare its divergence structure to the corresponding
  divergences in the dual metric from Eq. (\ref{eqCFTDualFIM}), which we write here
  again
  \begin{equation}\label{eqFisherCFT} G_{\lambda\lambda}^{(CFT)}=
  C \epsilon^{1-a},\quad a>1, \end{equation}
  with $C$ given in (\ref{eqCFTDualFIMC}). We now show that there exists
  a certain well defined limit at which HFIM asymptotically matches
  $G_{\lambda\lambda}^{(CFT)}$ on the boundary of the spacetime.
	
  \subsection{Dominant $\epsilon$}
	
  Let us first consider $\epsilon$ approaching zero sufficiently faster than
  $\mu\sim \tilde \varepsilon^2$. In this case, the dominant regulator is given
  by $\epsilon$, thus one can use the asymptotic expansion
  \begin{equation}\label{key} \Gamma \left( {a,z} \right)\sim{z^{a - 1}}{e^{
  - z}}\sum\limits_{k = 0}^\infty  {\frac{{\Gamma (a)}}{{\Gamma (a - k)}}{z^{
  - k}}} ,\quad |z| \to \infty, \end{equation}
  which defines the explicit divergence structure of the $b_i$ terms. On the
  other hand, the asymptotic expansion for the $a_i$ terms is governed by 
  \begin{equation}\label{eqZeroAsymptG} {\Gamma \left( {a,z} \right)}=
  \Gamma(a),\quad z \to 0,\quad a>0.  \end{equation}
  Therefore, Eq. (\ref{eqFisherFull}) acquires the following asymptotic form
  near $r=0$:
  \begin{align}\label{eqFisherH1} \nonumber &\mathcal{G}_{\lambda\lambda}
    =\left( a_0 + a_6 \Gamma^2 (a) \right) \frac{T\, \epsilon^{a-1}}{\mu^ a}
    \,-\, a_0 \left( \frac{\epsilon}{\mu}\right) ^{\!a} \,+\, a_1\, \Gamma
    (a-2) \frac{T^{2-a}\, e^{ - \frac{\mu }{T}}\, \epsilon
    ^{a-1}}{\mu}\\\nonumber &+ \left( a_2\, \Gamma^2 (a-2) + a_3\, \Gamma
  (2a-4) + a_4\, \Gamma (2a-2) + a_5\, \Gamma (2a-3) + a_7\, \Gamma^2 (a-1)
  \right) \left(\frac{\epsilon}{\mu}\right)^{\!a-1}\\\nonumber &+ b_1 e^{
-\frac{{2\mu }}{\epsilon}} \sum\limits_{k = 0}^\infty  {\frac{{\Gamma
(a\!-\!2)}}{{\Gamma (a\!-\!2\!-\!k)}} {\left(\frac{\epsilon}{\mu}\right)^{\!4
- a + k}}}  + b_2 e^{ - \frac{{2\mu }}{\epsilon }} \sum\limits_{k = 0}^\infty
{\sum\limits_{p = 0}^\infty  {\frac{{{\Gamma^2}(a - 2)}}{{\Gamma
(a\!-\!2\!-\!k)\, \Gamma (a\!-\!2\!-\!p)}}} {\left(\frac{\epsilon}{\mu}\right)
^{\!5 - a + k + p}}} \\\nonumber &+ b_3 e^{ -\frac{{2\mu }}{\epsilon }}
\sum\limits_{k = 0}^\infty  {\frac{{\Gamma (2a - 4)}}{{\Gamma
(2a\!-\!4\!-\!k)}}{2^{2a-5 - k}}{\left(\frac{\epsilon}{\mu}\right)^{\!4
- a + k}}}  + b_4  e^{ -\frac{{2\mu }}{\epsilon}} \sum\limits_{k = 0}^\infty
{\frac{{\Gamma (2a - 2)}}{{\Gamma (2a\!-\!2\!-\!k)}}{2^{2a-3
- k}}{\left(\frac{\epsilon}{\mu}\right) ^{\!2 - a + k}}} \\\nonumber &+ b_5 e^{
- \frac{{2\mu }}{\epsilon }} \sum\limits_{k = 0}^\infty  {\frac{{\Gamma (2a
- 3)}}{{\Gamma (2a\!-\!3\!-\!k)}} {2^{2a-4
- k}}{\left(\frac{\epsilon}{\mu}\right) ^{\!3 - a + k}}} + b_6 e^{
- \frac{{2\mu }}{\epsilon}} \sum\limits_{k = 0}^\infty  {\sum\limits_{p
= 0}^\infty  {\frac{{{\Gamma ^2}(a)}}{{\Gamma (a\!-\!k)\Gamma (a\!-\!p)}}}
{\left(\frac{\epsilon^{}}{\mu^{}}\right)^{\!2-a + k + p}}} \\ & + b_7 e^{ -2
\frac{\mu }{\epsilon }} \sum\limits_{k = 0}^\infty  {\sum\limits_{p = 0}^\infty
{\frac{{{\Gamma ^2}(a - 1)}}{{\Gamma (a\!-\!1\!-\!k)\Gamma (a\!-\!1\!-\!p)}}}
{\left(\frac{\epsilon^{}}{\mu^{}}\right)^{\!3-a + k + p}}}  .  \end{align}
\normalsize Note that the equals sign in Eq. \eqref{eqFisherH1} has to be understood as
  an asymptotic expansion. Obviously the divergence
  structure of any term depends on the range spanned by the parameter $a$. For
  example, the first term has a divergence structure $ T\,\epsilon ^{a
  - 1}\mu^{-a}$ for $T\to \infty, \mu\to 0, \epsilon \to 0$.  Therefore, one
  can take  $T\,\epsilon^{a - 1}\mu^{-a}= k=const$ for $a>1$, thus it can be
  considered regular. This leads to a vanishing term $T^{2-a} e^{ - \frac{\mu
  }{T}} \epsilon^{a - 1} \mu^{-1} \to 0$. The terms  with $\epsilon^{a} \mu^{-
a}$ and  $\epsilon^{a - 1} \mu ^{1 - a}$ also vanish for $a>1$, because
$\epsilon\to 0$ is dominant.  The other terms look complicated, but fortunately
they all have suppressing weight factors of $e^{-\frac{2\mu}{\epsilon}}\to 0$,
thus they all vanish for $a>1$. This analysis suggests that one can recover
the divergence structure of the dual CFT quantum information metric
(\ref{eqFisherCFT}) from (\ref{eqFisherH1}) for $a>1$, namely
  \begin{equation}\label{eqKey} {G_{\lambda \lambda }^{Bulk}} =\left( a_0 + a_6
  \Gamma^2(a) \right) k\, \epsilon^{1-a} =K \epsilon^{1-a}, \quad a>1,
\end{equation}
  where $K=const$. A similar result is valid also for finite $T$,
  \begin{equation}\label{eqKey1} {G_{\lambda \lambda }^{Bulk}} =\left( a_0
  + a_6 \Gamma^2 (a) \right) T k_1\, \epsilon^{1-a} =K_1\, \epsilon^{1-a},
\quad a>1, \end{equation}
  where $k_1= \epsilon^{a-1} \mu^{-a} =const$.  In order to complete the
  analysis one can further require the normalization constants $K$ from
  (\ref{eqKey}) and $C$ from (\ref{eqCFTDualFIMC}) to coincide, which can be
  used to fix $k$, namely
  \begin{equation}\label{eqNormConst} k = \kappa \frac{{(2a + d + 2)\Gamma (a
  - 1)\, {e^{\frac{{i\pi }}{2}(d - a - 1)}}}}{{{2^{a - 1}}{L^{d }}(2 - 2a
  + d)\left( {\Gamma (a)\Gamma (a + 1) - \Gamma (2a + 1){B_{1/2}}(a,a)}
  \right)}}.  
\end{equation}
	
Insisting on holographic duality on both sides of the correspondence forces us to equate the coefficients in front of the leading singularities of both theories. Interestingly, being in perturbative regime in the bulk and staying close to the boundary the parameters characterizing gravity side shows the behaviour of the theory in the limiting case. 
  
\subsection{Dominant $\tilde \varepsilon$}
	
When $\tilde \varepsilon$ approaches zero sufficiently faster than
$\epsilon$, the dominant regulator is given by $\mu$. In this case, the
second argument of the incomplete gamma function goes to zero, thus we can
use (\ref{eqZeroAsymptG}) for all terms. Hence, one finds
  \begin{align}\label{} \nonumber &\mathcal{G}_{\lambda\lambda} = \left( a_0
    + a_6 \Gamma^2 (a) \right)\frac{T \epsilon^{a - 1} }{\mu^a} - a_0
    \left(\frac{\epsilon}{\mu}\right)^{\!a} + a_1 \Gamma (a-2) \frac{T^{2
    - a}\, e^{ - \frac{\mu }{T}}\, \epsilon ^{a-1}}{\mu}\\\nonumber &+
    \left(a_2 \Gamma^2 (a - 2) + a_3 \Gamma (2a-4) + a_4 \Gamma (2a-2) + a_5
    \Gamma (2a-3) + a_7 \Gamma^2 (a-1) \right)
    \left(\frac{\epsilon}{\mu}\right)^{\!a - 1} \\\nonumber &+ \left( b_2
    \Gamma^2 (a-2) + b_3 \Gamma (2a-4) + b_4 \Gamma (2a-2) + b_5 \Gamma (2a-3)
    + b_7 \Gamma^2(a-1) \right) \left(\frac{\epsilon}{\mu}\right)^{\!a - 1}\\
                                                            &+ b_1 \Gamma (a-2)
                                                            \frac{\epsilon}{\mu}
                                                            + b_6 \Gamma^2 (a)
                                                            \left(\frac{\epsilon}{\mu}\right)^{\!a}.  
\end{align}
  Due to the fact that $a_i=-b_i$, the terms with $a_2,b_2, a_3, b_3, a_4,
  b_4,a_5, b_5$ and $a_7, b_7$ cancel each other out, leaving us only with
  \small \begin{align}\label{eqFisherH2A} \mathcal{G}_{\lambda\lambda} =\left(
    a_0 + a_6 \Gamma^2 (a) \right) \frac{T \epsilon ^{a - 1}}{\mu^a}
    - \left(a_0 + a_6 \Gamma^2 (a) \right)
    \left(\frac{\epsilon}{\mu}\right)^{\!a} + a_1 \Gamma (a-2) \frac{ T^{2-a}
    \epsilon^{a-1} }{\mu} - a_1 \Gamma (a-2) \frac{\epsilon}{\mu}
  .  \end{align}
  \normalsize
  If $T\to \infty$,  all terms in the metric are divergent for $1<a\leq 2$, while
  for $a>2$ only the third term can be considered regular. For finite $T$ all
  terms remain divergent for $a>1$. This suggests that when $\mu$ dominates the
  divergent structure makes it impossible to represent HFIM in the form given by
  Eq.(\ref{eqFisherCFT}), thus one cannot obtain the dual QFIM from the bulk.  
	
  \subsection{The case $\epsilon\sim\tilde \varepsilon$}
	
  In this special case we  have effectively one divergence, namely $0<\eta\ll
  1$. Hence, the asymptotic behaviour is the same as in the previous case, but
  with $\mu\sim \eta^2$ and $\epsilon\sim \eta$ replaced, thus
  \begin{equation}\label{eqFisherH2} \mathcal{G}_{\lambda\lambda}
    =\frac{2^a}{M^a}\! \left(a_0 + a_6 \Gamma^2 (a) \right) \frac{T-\eta}{\eta
    ^{a+1}}  \,+\,  \frac{2a_1}{M} \Gamma (a-2) \left( \frac{T^{2
  - a}}{\eta^{3-a}} -  \frac{1}{\eta} \right).  \end{equation}
  Taking $T\to \infty$ and $ \eta\to 0$,  all terms are divergent for $1<a\leq
  2$. If $2<a< 3$, only the term $T^{2-a}\eta^{a-3}$ can be fixed to a constant.
  If $a\geq 3$, the third term vanishes. When $1<a<3$ and $T$ is considered
  finite, all terms diverge. If $a\geq 3$, the third terms is regular or vanishes
  for finite $T$. Therefore, it is impossible to reproduce the divergent
  structure of the dual CFT metric from the bulk HFIM.  \\
	
  Let us briefly summarize the results for HFIM near the boundary $r=0$. When
  approaching asymptotically the boundary, primarily with $\epsilon\to 0$, we
  have been able to fully match the structure of the dual CFT QFIM by the
  structure of the bulk HFIM. On the other hand, for predominant
  $\tilde{\varepsilon}\to 0$, HFIM is intrinsically divergent on the boundary
  with a structure that does not coincide with the dual CFT QFIM. Also, for the
  particular case $\epsilon\sim\tilde{\varepsilon}\to 0$, the dual QFIM can not
  be reproduced by the bulk HFIM.
	
  \section{Analysis of the holographic Fisher information metric}\label{sec6}
	
  In this section we are going to study the full divergence structure of HFIM
  in the bulk along the holographic coordinate $r$. This requires considering
  HFIM from Eq. (\ref{eqFisherBulk}).
	
  \subsection{HFIM near the boundary $r\to 0$}
	
  In Section \ref{sec5} we have shown that one can reproduce the divergence
  structure and even match the normalization constant of the boundary CFT QFIM
  from the bulk HFIM under certain conditions. This was done by explicitly taking
  out a factor of $\epsilon^{a-1}$ and considering the divergence structure for
  the rest of the metric. Below we will restore this factor and consider the
  full HFIM near the boundary $r=0$. Once again, one has several cases.  \\

  \textbf{Dominant $\epsilon$}. In this case, the bulk Fisher metric
  (\ref{eqFisherBulk}) has the following asymptotic form
  \begin{align}\label{eqFisherHB2} \nonumber &{G_{\lambda \lambda }^{Bulk}}
    =\left( a_0 + a_6\, \Gamma^2 (a) \right) \frac{T}{\mu^a} \,-\, a_0
    \frac{\epsilon}{\mu^a} \,+\, a_1\, \Gamma (a-2)  \frac{T^{2 - a} e^{
    - \frac{\mu}{T}}}{\mu}\\\nonumber &+ \left( a_2\Gamma^2 (a-2) + a_3 \Gamma
  (2a-4) + a_4 \Gamma (2a-2) + a_5 \Gamma (2a-3) + a_7 \Gamma^2 (a-1) \right)
    \frac{1}{\mu^{a - 1}} \\\nonumber &+ b_1 e^{ - \frac{{2\mu }}{\epsilon }}
    \sum\limits_{k = 0}^\infty  \frac{\Gamma (a-2)\, \epsilon^{1-a}} {\Gamma
    (a-2-k)} \left(\frac{\epsilon}{\mu}\right) ^{\!4 - a + k}  + b_2 e^{
  - \frac{2\mu }{\epsilon }} \sum\limits_{k = 0}^\infty  \sum\limits_{p
= 0}^\infty  \frac{\Gamma^2 (a-2)\, \epsilon^{1-a}} {\Gamma (a\!-\!2\!-\!k)\,
\Gamma (a\!-\!2\!-\!p)} \left(\frac{\epsilon}{\mu}\right) ^{\!5 - a + k + p}
\\\nonumber &+ b_3 e^{ - \frac{{2\mu }}{\epsilon }} \sum\limits_{k = 0}^\infty
\frac{\Gamma (2a-4)\,2^{2a-5-k}} {\Gamma (2a-4-k)}\, \epsilon^{1-a}
\left(\frac{\epsilon}{\mu}\right) ^{\!4 - a + k}  + b_4 e^{ - \frac{{2\mu
}}{\epsilon}} \sum\limits_{k = 0}^\infty  \frac{\Gamma (2a-2)\, 2^{2a-3-k}}
{\Gamma (2a-2-k)}\, \epsilon^{1-a} \left(\frac{\epsilon}{\mu}\right) ^{\!2
- a + k} \\\nonumber &+ b_5 e^{ - \frac{{2\mu }}{\epsilon }} \sum\limits_{k
= 0}^\infty  \frac{\Gamma (2a-3)\, 2^{2a-4-k}} {\Gamma (2a-3-k)}\,
\epsilon^{1-a} \left(\frac{\epsilon}{\mu}\right) ^{\!3 - a + k} + b_6 e^{
- \frac{{2\mu }}{\epsilon}} \sum\limits_{k = 0}^\infty  \sum\limits_{p
= 0}^\infty  \frac{\Gamma^2 (a)\, \epsilon^{1-a}} {\Gamma (a - k)\, \Gamma (a
- p)} \left(\frac{\epsilon}{\mu}\right)^{\!2-a + k + p} \\ & + b_7 e^{ -2
\frac{\mu }{\epsilon }} \sum\limits_{k = 0}^\infty  \sum\limits_{p = 0}^\infty
\frac{\Gamma^2 (a-1)\, \epsilon^{1-a}} {\Gamma (a-1-k)\, \Gamma (a-1-p)}
\left(\frac{\epsilon}{\mu}\right)^{\!3-a + k + p}  .  \end{align} \normalsize
By taking out a factor of $\epsilon^{a-1}$ one returns to the HFIM from
(\ref{eqFisherH1}), which was used to match the divergence structure of the
dual QFIM on the boundary. On the other hand, when $T\to \infty$ and
$\epsilon\to 0$ being dominant,  all the $b_i$ terms in (\ref{eqFisherHB2}) are
suppressed due to the exponential weight factors. Moreover, the second term
with $a_0$ also vanishes, thus HFIM reduces to
  \begin{align}\label{eqFisherHB3} \nonumber {G_{\lambda \lambda
    }^{Bulk}}&=\left( a_0 + a_6\, \Gamma^2 (a) \right) \frac{T}{\mu^a}  \,+\,
    a_1\Gamma (a-2) \frac{T^{2 - a}}{\mu}\\ &+ \left( a_2 \Gamma^2 (a-2) + a_3
    \Gamma (2a-4) + a_4 \Gamma (2a-2) + a_5 \Gamma (2a-3) + a_7 \Gamma^2 (a-1)
  \right) \frac{1}{\mu^{a - 1}}.  \end{align}
  One notes that the first and third terms are intrinsically divergent for
  $a>1$, while the second term is divergent only for $1<a\leq 2$. The latter
  can be regular for $a>2$. When $T$ is finite -- all therms are divergent.  \\
	
  \textbf{Dominant $\tilde \varepsilon$}. When $\tilde \varepsilon$ goes to
  zero sufficiently faster than $\epsilon$, one finds
  \begin{align}\label{eqFisherH2B} {G_{\lambda \lambda }^{Bulk}} &=\left( a_0
  + a_6 \Gamma^2 (a) \right) \frac{T - \epsilon}{\mu^a} \,+\, a_1 \Gamma
(a-2)\, \frac{T^{2-a} - \epsilon^{2-a}}{\mu} .  \end{align}
  Taking $T\to \infty$,  all terms diverge for $a>1$, except for the term with
  $ T^{2-a}$, which is divergent only for $1<a\leq 2$ and can be regular for
  $a>2$. When $T$ is considered finite, all terms are divergent for $a>1$.  \\
	
  \textbf{The case $\epsilon\sim\tilde \varepsilon$}. When neither of the
  regulators dominates one effectively has $\epsilon=\tilde \varepsilon=\eta$,
  hence
  \begin{align} {G_{\lambda \lambda }^{Bulk}}=\frac{2^a}{M^a}\! \left( a_0
  + a_6\, \Gamma^2 (a) \right) \frac{T-\eta}{\eta ^{2a}}  \,+\,  \frac{2a_1}{M}
\Gamma (a-2)\, \frac{T^{2 - a}-\eta^{2-a}}{\eta^2}.  \end{align}
  In the limit $T\to\infty$, all term are divergent, except for term with
  $ T^{2-a}$, which is divergent only for $1<a\leq 2$, and regular for $a>2$.
  For finite $T$ -- all terms diverge for $a>1$.

  \subsection{HFIM in the bulk} In the bulk of Schr\"odinger spacetime the
  holographic coordinate $r$ is finite and the divergence structure of the
  metric is managed by $\epsilon$ and $T$. The following cases are relevant.
  \\
	
  \textbf{The case $\mu\to const\neq 0, \,T\to \infty$}. When we consider HFIM
  in the bulk $\mu\sim r^2$ is finite. If one takes $T\to \infty$ with
  $\epsilon\to 0$, Eq. \eqref{eqFisherBulk} becomes 
  \begin{align}\label{eqFisherBulk1} \nonumber {G_{\lambda \lambda }^{Bulk}}
  &=\left( a_0 + a_6\, \Gamma^2 (a) \right) \frac{T}{\mu^a}  \,+\, a_1\Gamma
    (a-2) \frac{T^{2 - a}}{\mu}\\ &+ \left( a_2 \Gamma^2 (a-2) + a_3 \Gamma
    (2a-4) + a_4 \Gamma (2a-2) + a_5 \Gamma (2a-3) + a_7 \Gamma^2 (a-1) \right)
    \frac{1}{\mu^{a - 1}}.  \end{align}
  One notes that the divergence structure of the information metric is governed
  fully by $T$. For $1<a\leq 2$ both terms are divergent, while for $a>2$ only
  the first term is divergent.  \\
	
  \textbf{The case $\mu\to const\neq 0, \,T\to const$}. With both $\mu$ and $T$
  finite the holographic Fisher information metric is completely regular for
  $a>1$, although it has the same form as in Eq. (\ref{eqFisherHB3}). In other
  words, the information space now is non-singular.
	
  \subsection{HFIM near $r\to\infty$} Finally, we are going to consider HFIM on
  the second boundary at $r\to \infty$.  \\
	
  \textbf{The case $\mu\to \infty, \,T\to const$}.  Here, the bulk HFIM
  \eqref{eqFisherBulk} vanishes trivially. One way to interpret this result is
  that one cannot discern between  deformed theory with coupling
  $\delta\lambda$ and the undeformed theory at $r\to\infty$. In other words,
  the Fisher distance between both theories in the space of coupling constants
  is zero.  \\
	
  \textbf{The case} $\mu\to \infty$ (dominant), $T\to \infty$. When we are
  approaching asymptotically the boundary at $r\to\infty$ with predominant
  $\mu\to\infty$, the holographic Fisher information metric
  \eqref{eqFisherBulk} also vanishes trivially.  \\
	
  \textbf{The case} $\mu\to \infty$, $T\to \infty$ (dominant). We have only two
  relevant terms
	
  \begin{align} {G_{\lambda \lambda }^{Bulk}} =( {a_0}+ {a_6}{\Gamma ^2}\left(
  {a} \right))\frac{T}{\mu ^{ a}}  + {a_1}\Gamma \left( {a - 2} \right)
\frac{T^{2 - a}}{\mu}, \end{align}
  which can be considered divergent for $1<a<2$. When $a>2$ she second term
  vanishes and we are left only with  
  \begin{equation}\label{eqNonSinga} {G_{\lambda \lambda }^{Bulk}} =( {a_0}+
  {a_6}{\Gamma ^2}\left( {a} \right))\frac{T}{\mu ^{ a}} , \end{equation}
  which is regular if one considers $T\mu^{-a}=const$.  \\
	
  In summary, we have studied the divergence structure of the HFIM in the bulk
  of the entire Schr\"odinger spacetime, together with its boundaries, along
  the holographic direction $r$. We have shown that depending on how we
  approach the given sectors, different divergence structures
    arise, managed
  by the relevant regulators  $\tilde{\varepsilon}$, $T$ or $\epsilon$. In many
  of the considered cases HFIM is intrinsically divergent. However, we have
  found two  cases, in which the bulk HFIM is regular and finite, thus the
  information space over the couplings of the theory is now a well-defined
  non-singular Riemannian manifold. Finally, we have encountered two cases at
  $r\to\infty$, where the HFIM is zero, which suggest that we cannot discern
  between a deformed theory with coupling $\delta\lambda$ and the undeformed
  one. 

  \section{Conclusion}\label{sec7} In this study we have investigated the
  properties of the Fisher information metric on both sides of the duality
  between non-relativistic dipole gauge theory and bulk string theory in
  Schr\"odinger spacetime\footnote{In this case, the bulk theory is
  $(d+3)$-dimensional Schr\"odinger spacetime and the dual gauge theory lives
on the $(d+1)$-dimensional boundary.  }. Our work extends the scope of
information theory from the original AdS/CFT correspondence to non-relativistic
holography. To the best of our knowledge this is the first attempt of application of information theory to Schr\"odinger/Dipole holography. The setup consists of a marginally deformed CFT by inserting
a primary operator at Euclidean time $\tau = 0$. Holographically, this
corresponds to a massive scalar field probing the Schr\"odinger background
geometry on the string side, where backreaction on the original spacetime has
been considered negligible. This is reflected by the perturbative technique
used to calculate the holographic Fisher information metric as shown in
\cite{Trivella:2016brw}.
	
  On the gauge side, the quantum Fisher information metric on the space of
  coupling constants has been computed via the two-point correlation function
  between relevant deformation operators at different times $\tau_1$ and
  $\tau_2$. The computation has been done in the ground states of the deformed
  and the undeformed theories. The final expression \eqref{eqCFTDualFIM} for
  the QFIM is independent of any spacetime coordinates and represents the
  information content difference between the originally
  undeformed CFT and the
  one generated after turning on the deformation. In this case, the divergence
  structure of the QFIM consists only of one temporal regulator $\epsilon\to
  0$, coming from the fact that the CFT Lagrangian changes discontinuously at
  $\tau=0$, which introduces an UV divergence.
	
  On the string side, we have used the perturbation method suggested in
  \cite{Trivella:2016brw}, which is valid in the large $N$ limit of the theory,
  in order  to compute the dual HFIM given in Eq. \eqref{eqFisherBulk}. This
  metric exhibits a richer divergence structure by introducing not one, but
  three different regulators -- two temporal $\epsilon\to 0$ and $T\to \infty$,
  and one spacial regulator $\tilde{\varepsilon}$ along the holographic
  direction $r$. As shown on Fig. \ref{fig1}, we can divide the  Schr\"odinger
  background along the coordinate $r$ in three sectors. One of them includes
  the boundary at $r=0$, where the dual CFT lives. The second one includes the
  bulk of spacetime, and the third sector is the limit $r\to\infty$. This
  naturally introduces several important cases.
   
 The most important case for us is to
  compare the asymptotic divergent structure of HFIM to QFIM near the boundary 
  at $r=0$. The latter turn out to be a nontrivial task due to the competing divergences in the HFIM. Our analysis showed that when approaching asymptotically the boundary at $r=0$, primarily with
  $\epsilon\to 0$, we have uncovered a full match between the leading divergences
  \eqref{eqKey}  of HFIM and the dual CFT QFIM. This result means that one can
  reconstruct the entire QFIM from the bulk HFIM close to the boundary. Furthermore, for predominant
  $\tilde{\varepsilon}\to 0$, HFIM is intrinsically divergent on the boundary
  with a structure that does not coincide with the dual QFIM from
  \eqref{eqCFTDualFIM}. This is also true for the particular case of
  $\epsilon\sim\tilde{\varepsilon}\to 0$. In summary, we showed that there exist only one asymptotic limit of HFIM, which leads to a consistent Schr\"odinger/Dipole theory duality. 
	
  We have further studied the divergence structure of HFIM in the bulk of the
  entire Schr\"odinger spacetime together with its boundaries. Depending on how
  we approach different sectors, we have found specific divergence structures
  managed by the relevant regulators  $\tilde{\varepsilon}$, $T$ or $\epsilon$.
  In many of the analysed cases HFIM is intrinsically divergent. However, there
  appeared two  cases, where the bulk HFIM is regular and finite, thus the
  information space over the coupling constants of the theory is a non-singular
  Riemannian manifold. Finally, we have encountered two cases at $r\to\infty$,
  where HFIM vanishes, thus one cannot discern between a deformed theory with
  coupling $\delta\lambda$ and the undeformed one.
	
  Although our study covers the computation of the holographic Fisher
  information metric and its dual quantum counterpart for non-relativistic
  holographic models in Schr\"odinger spacetime, there are still many
  unexplored research directions. For example, it would be interesting to
  consider key information-theoretic aspects of non-relativistic holography in
  the MERA and cMERA (continuous multi-scale entanglement renormalization
  ansatz) approach to the gauge/gravity correspondence \cite{Miyaji:2015yva},
  where HFIM in the bulk plays an important role. Furthermore, the HFIM for
  systems at finite temperature, which traditionally involves the presence of
  black holes, is also interesting to consider. Another interesting direction of
  investigation is to apply other relevant or irrelevant deformations on the
  theory with  multiple operators. Examples of such deformations include
  $T\bar{T}, JT, J\bar{T}$, etc. It could prove fruitful to consider models in
  different non-relativistic holographic solutions such as Lifshitz
  spacetimes\footnote{Such investigations for holographic entanglement entropy,
    mutual information and entanglement of purification over holographic
  Lifshitz theory have been already initiated in \cite{Gong:2020pse}.} or other
  similar backgrounds. Finally, it could be interesting to investigate the
  relation between the Fisher information metric
 and quantum complexity for such models.

  \subsection*{Acknowledgements} The authors would like to thank Anastasia
  Golubtsova and Peter Ivanov for insightful discussions. The support given by
  the Bulgarian NSF grants N28/5 and DN-18/1 is gratefully acknowledged. This
  work is also partially supported by the Program “JINR– Bulgaria” at the
  Bulgarian Nuclear Regulatory Agency. M. R.and I. I. acknowledge the support by the Bulgarian national program “Young Scientists
  and Postdoctoral Research Fellows” 2021.
	
  \begin{appendix}
		
  \section{A short note on how to generate Schr\"odinger spacetimes and their
  dual CFT theories} \label{app0}
		
  The symmetry of the free Schr\"odinger equation 
    \begin{equation} \frac{\partial^2}{\partial\vec{r}^{\,2}}\phi -2im
    \frac{\partial}{\partial t}\phi =0, \end{equation}
    is the so called Schr\"odinger group. In $n$ dimensional spacetime the
    group consists of spatial translations indicated by $\vec A$, rotations
    given by the matrix $\Omega$, and Galilean boosts with velocity $\vec v$,
    \begin{equation} t\:\to\: t'=\frac{at+b}{ct+d}, \qquad \vec{r}\:\to\:
    \vec{r}\,'=\frac{\Omega\vec{r} + \vec{v}t + \vec{A}}{ct+d},\quad ad-bc=1.
  \end{equation}
    In addition, one has dilatation, where time and space scale differently
    \begin{equation}\label{key} t\to \lambda^2 t,\quad\vec{r}\to\lambda
    \vec{r}, \end{equation}
    and one additional special conformal transformation
    \begin{equation}\label{key} t\to\frac{t}{1+\lambda t},\quad \vec r \to
    \frac{\vec r}{1+\lambda t}.  \end{equation}

    From group theory point of view, the Schr\"odinger group can be thought of
    as a non-relativistic analogue of the conformal group. In fact, the
    Schr\"odinger group can be embedded into the relativistic conformal group
    $SO(2,n+2)$ in $n+1$ dimensions \cite{Son:2008ye, Balasubramanian:2008dm,
    barut-niederer}, as well as a particular contraction of the conformal
    group. 
		
    For purposes of the holographic correspondence it is important to consider
    spaces with the Schr\"odinger group being the maximal group of
    isometries\footnote{For a detailed group-theoretical perspective on
    non-relativistic holography see \cite{ Dobrev:2013kha}.}. Such spaces are
    called Schr\"odinger spaces.  There is a specific way to obtain the
    spacetime geometry equipped with this symmetry via TsT
    (T-duality-shift-T-duality) transformations. 
		
    Our starting point is the AdS metric, which is invariant under the whole
    conformal group and  deform it to reduce the symmetry down to the
    Schr\"odinger group. 
		
    Being a particular case of the so called Drinfel'd-Reshetikhin twist, the
    TsT procedure has been used for generating many backgrounds keeping partial
    or full integrability of the system. Specific point in generating
    Schr\"odinger backgrounds via TsT transformations is to include one of the
    light-cone variables. This particular deformation is also known as the
    null-Melvin twist. This procedure ca be implemented by the following key
    steps:
		
    \begin{itemize} \item Represent the theory in light-cone coordinates and
      identify a Killing direction, say $\psi$, \item Perform a T-duality along
      the chosen Killing direction $\psi$, \item Boost the geometry in the
      Killing direction by $\hat{\mu}$,  i.e. $x^- \to x^- - \hat{\mu}
      \tilde{\psi}$, where $\tilde{\psi}$ is the T-dualized coordinate $\psi$,
\item Finally, perform a T-duality back to IIA/IIB along $\tilde \psi$.
\end{itemize}
		
    In order to accomplish the desired result consider general background in
    the form $AdS_n\times X^m$,
    \begin{equation} ds^2=\ell^2\,\frac{2dx^+dx^-+dx^idx_i
    +dz^2}{z^2}+ds^2_{X^m}, \end{equation}
 Now we perform a null
     Melvin twist along a Killing vector $\mathcal{K}$ on $X^m$. The result is
    \begin{equation} ds^2=ds^2_{Schr_n}+ds^2_{X^m}, \end{equation}
		where the Schr\"odinger metric yields
    \begin{equation}
    ds^2_{Schr_n}=-\frac{\Omega}{z^4}+\frac{1}{z^2}(2dx^{+}dx^{-}+d\vec
  x^2+dz^2), \quad \Omega= \left\| \mathcal{K}\right\|
^2=\star(\mathcal{K} \wedge\star\mathcal{K}). \end{equation}
   Here the Hodge star operator $\star$ is taken with respect to the metric on $X^m$. It is clear that $\Omega$ is
    non-negative being a square length of a Killing vector\footnote{Together
      with the metric the procedure also generates a non-zero $B$-field of the
      form
      \begin{equation*} B_{(2)}=\frac{1}{z^2}\mathcal{K}\wedge dx^+.
      \end{equation*}
    }. For particular examples see \cite{Golubtsova:2020mjn,
    Golubtsova:2020fpm, Matsumoto:2015uja, Kameyama:2015ufa, Georgiou:2017pvi}\footnote{The original Schr\"odinger background
    (\ref{eqSchrMetric}), used in this paper, was first derived in
    \cite{Son:2008ye, Balasubramanian:2008dm}.}.
		
    An important remark is that in order to make holographic sense of these
    solutions one has to impose some conditions. In particular for these to be
    holographic duals to non-relativistic field theories the light-cone
    coordinate $x^-$ should be periodic, $x^-\sim {x^-} +2\pi r_{x^-}$
    \cite{Son:2008ye,Balasubramanian:2008dm,Adams:2008wt}. The momentum along
    this compact direction is quantized in units of the inverse radius
    $r_{x^-}^{-1}$.

    The dual theory is conjectured to be a specific field theory, namely dipole
    field theory. The TsT transformation, which produces the bulk theory
    corresponds to a particular deformation (Drinfel'd-Reshetikhin twist) on
    field theory side. It translates to the dual theory as a star product. When
    the directions involved in TsT are transverse to the stack of branes from
    which geometry descends, it produces a twist in the field theory having
    ordinary product. If however, one of the directions is along the branes,
    then the star product is non-trivial and the theory becomes dipole one
  \begin{equation}\label{key} (\Phi_1\star\Phi_2)(x)=
  \Phi_1(x+L_1)\Phi_2(x-L_2), \end{equation}
    where $L=L_1+L_2$ is the dipole length associate with R-charges. The big
    advantage of holographic model with Schr\"odinger symmetry is that they
    could be integrable. Proving that in the case of $Schr_5\times S^5$ for
    example, the authors of \cite{Guica:2017jmq} were able to map the composite
    operators of monomial form to a spin chain. To study field theory side one
    must just to replace the ordinary product with the star one
  \begin{equation}\label{key} \mathcal{O}=
  \operatorname{tr}(\Phi_1\star\Phi_2\star\cdots).  \end{equation}
    Here, one can choose to work either using Seiberg-Witten map or working
    directly with the star product. A nice analysis has been presented in
    \cite{Guica:2017jmq}.

\section{Derivation of the CFT quantum geometric tensor} \label{appB1}
 	
Let $\ket{\varphi}$ be a generic state,
inserted at
$\tau=0$. The overlap between the original ground state $\ket{\psi_0}$ at
$\tau\to-\infty$ and this new state at $\tau\to 0$ is given by
\begin{equation}\label{key}
	\braket{\tilde\varphi|\psi_0}=\frac{1}{\sqrt{Z_0}}\int\limits^{\tilde
		\varphi}\mathcal{D}\varphi \,e^{-\int\limits_{-\infty}^0 d\tau\int d^d
		x \mathcal{L}_0}, \end{equation}
where $\tilde\varphi=\varphi(\tau=0)$ and the partition function of the
initial undeformed theory is defined by

\begin{equation}\label{key} Z_0=\int\mathcal{D}\varphi\,
	e^{-\int\limits_{-\infty}^\infty d\tau\int d^d x \mathcal{L}_0}.
\end{equation}

In a similar fashion, one can consider the evolution from $\tau= 0$, where
$\ket{\tilde\varphi}$ is inserted, to $\tau\to\infty$, where we are placing
the perturbed state $\ket{\psi_1}$,
\begin{align}\label{key}
	\braket{\psi_1|\tilde\varphi}=\frac{1}{\sqrt{Z_1}}\int\limits_{\tilde
		\varphi}\mathcal{D}\varphi \,e^{-\int\limits_{0}^\infty d\tau\int d^d
		x \mathcal{L}_1}=\frac{1}{\sqrt{Z_1}}\int\limits_{\tilde
		\varphi}\mathcal{D}\varphi \,e^{-\int\limits_{0}^\infty d\tau\int d^d x(
		\mathcal{L}_0+\delta\lambda^a\mathcal{O}_a)}.  \end{align}
As usually,
\begin{equation} Z_1=\int\mathcal{D}\varphi
	\,e^{-\int\limits_{-\infty}^\infty d\tau\int d^d x(
		\mathcal{L}_0+\delta\lambda^a\mathcal{O}_a)} \end{equation}
is the partition function of the deformed theory.  Now, the overlap between
both states in the above notations can be formally written as
\begin{equation}\label{eqPathIntQIMa}
	\braket{\psi_1|\psi_0}=\int\limits_{\varphi(\tau=0)=\tilde\varphi}\mathcal D\varphi \braket{\psi_1| \varphi}\braket{
		\varphi|\psi_0}=\frac{1}{\sqrt{Z_0 Z_1}} \int\mathcal{D}\varphi \,e^{-\int
		d^d x \left[\int\limits_{-\infty}^0 d\tau\mathcal{L}_0+\int\limits_{0}^\infty
		d\tau( \mathcal{L}_0+\delta\lambda^a\mathcal{O}_a)\right]}.  
\end{equation}
Let us write this overlap in a more convenient way
\cite{Alvarez-Jimenez:2019ytg}
\begin{equation}\label{key}
	\braket{\psi_1|\psi_0}=\frac{\left\langle\exp\left({-\int\limits_{0}^\infty
			d\tau\int \limits_{V_{\mathbb{R}^d}} d^d x \delta\lambda^a
			\mathcal{O}_a}\right)\right\rangle}{\sqrt{\frac{Z_1}{
				Z_0}}}=\frac{\left\langle\exp\left({-\int\limits_{0}^\infty d\tau\int
			\limits_{V_{\mathbb{R}^d}} d^d x \delta\lambda^a
			\mathcal{O}_a}\right)\right\rangle}{\sqrt{\left\langle\exp\left(-\int\limits_{-\infty}^\infty
			d\tau\int\limits_{V_{\mathbb{R}^d}}d^d x\delta\lambda^a\mathcal
			O_a\right)\right\rangle}}, \end{equation}
therefore, one has
\begin{equation}\label{key} |\langle \psi_1|\psi_0\rangle|^2
	= \frac{{\left\langle {\exp \left( { - \int\limits_0^\infty  d \tau
					\int\limits_{V_{\mathbb{R}^d}} {{d^d}} x\delta {\lambda ^a}{{\cal
							O}_a}(\tau )} \right)} \right\rangle \left\langle {\exp \left(
				{ - \int\limits_{ - \infty }^0 d \tau \int\limits_{V_{\mathbb{R}^d}}
					{{d^d}} x\delta {\lambda ^a}{{\cal O}_a}(\tau )} \right)} \right\rangle
	}}{{\left\langle {\exp \left( { - \int\limits_{ - \infty }^\infty  {d\tau
						\int\limits_{V_{\mathbb{R}^d}} {{d^d}} } x\delta {\lambda ^a}{{\cal
							O}_a(\tau)}} \right)} \right\rangle }}.  \end{equation}
We can now easily expand this expression in power series
\begin{align} \nonumber |\langle \psi_1|\psi_0\rangle|^2 &=
	1 + \frac{1}{2}\int\limits_{V_{\mathbb{R}^d}} {{d^d}x_1}
	\int\limits_{V_{\mathbb{R}^d}} {{d^d}x_2} \left[ {\int\limits_0^\infty
		{d{\tau _1}} \int\limits_0^\infty  {d{\tau _2}} \left\langle {{{\cal
					O}_a}({\tau _1}){{\cal O}_b}({\tau _2})} \right\rangle } \right.\\\nonumber
	&+ \int\limits_{ - \infty }^0 {d{\tau _1}} \int\limits_{ - \infty }^0 {d{\tau
			_2}} \left\langle {{{\cal O}_a}({\tau _1}){{\cal O}_b}({\tau _2})}
	\right\rangle - \int\limits_{ - \infty }^\infty  {d{\tau _1}} \int\limits_{
		- \infty }^\infty  {d{\tau _2}} \left\langle {{{\cal O}_a}({\tau _1}){{\cal
				O}_b}({\tau _2})} \right\rangle \\ &+ \left. 2{\int\limits_0^\infty  {d{\tau
				_1}} \int\limits_{ - \infty }^0 {d{\tau _2}} \left\langle {{{\cal O}_a}({\tau
				_1})} \right\rangle \left\langle {{{\cal O}_b}({\tau _2})} \right\rangle
	} \right]\delta {\lambda ^a}\delta {\lambda ^b} + {\cal O}(\delta {\lambda
		^3}).  \end{align}
Since we have time reversal symmetry of the correlator,
\begin{equation}\label{key}
	\braket{\mathcal{O}_a(-\tau_1)\mathcal{O}_b(-\tau_2)}=\braket{\mathcal{O}_a(\tau_1)\mathcal{O}_b(\tau_2)},
\end{equation}
and the fact that
\begin{equation}\label{key} \int\limits_{ - \infty }^\infty  {d{\tau _1}}
	\int\limits_{ - \infty }^\infty  {d{\tau _2}}  = \int\limits_{ - \infty }^0
	{d{\tau _1}} \int\limits_{ - \infty }^0 {d{\tau _2}}  + \int\limits_{
		- \infty }^0 {d{\tau _1}} \int\limits_0^\infty  {d{\tau _2}}
	+ \int\limits_0^\infty  {d{\tau _1}} \int\limits_{ - \infty }^0 {d{\tau _2}}
	+ \int\limits_0^\infty  {d{\tau _1}} \int\limits_0^\infty  {d{\tau _2}}
	, \end{equation}
one finds 
\begin{align}\label{key} |\langle \psi_1|\psi_0\rangle|^2=
	1 - \frac{1}{2}\int\limits_{V_{\mathbb{R}^d}} {{d^d}x_1}
	\int\limits_{V_{\mathbb{R}^d}} {{d^d}x_2} \int\limits_{ - \infty }^0 {d{\tau
			_1}} \int\limits_0^\infty  {d{\tau _2}} \left( {\left\langle {{{\cal
					O}_a}({\tau _1}){{\cal O}_b}({\tau _2})} \right\rangle  - \left\langle
		{{{\cal O}_a}({\tau _1})} \right\rangle \left\langle {{{\cal O}_b}({\tau
				_2})} \right\rangle } \right)\delta {\lambda ^a}\delta {\lambda ^b}.
\end{align}
Using the definition of the quantum fidelity,  we extract the expression for QIM
from Eq. (\ref{eqQIMCFTsidea}).

    \section{Computation of the dual CFT quantum Fisher metric}\label{appA} One
    has to compute the following integral expression
    \begin{equation} G_{\lambda \lambda }^{(CFT)} = \frac{i\Delta M^{\Delta -1}
      2^{-\Delta} e^{- \frac{i\pi \Delta}{2}} } {\pi^{d/2}\, \Gamma
    \left(\Delta  - \frac{d}{2} - 1\right)} \int\!\!d^d x_1 \int\!\!d^d x_2
    \int\limits_{ - \infty }^{ - \epsilon }\!\!d\tau_1
    \int\limits_\epsilon^\infty \!\!  d\tau_2 \, \frac{1}{(\tau_2
    - \tau_1)^{\Delta}} \, e^{i\frac{M(1+i \varepsilon)}{2} \frac{(\vec x_2
  - \vec x_1)^2}{\tau_2 - \tau_1}} , \end{equation}
    which is a Gaussian integral over $\vec x$ space, i.e.
    \begin{equation} {I_x} = \int\!\! {{d^{d }}{x_1}\int \!\! {{d^{d}}{x_2} \,
      {e^{i\frac{M(1+i \varepsilon)}{2}\frac{{{{({{\vec x}_2} - {{\vec
  x}_1})}^2}}}{{{\tau_2} - {\tau_1}}}}}} }  = {e^{\frac{{i\pi }}{2}(1
  - d/2)}}{(2\pi )^{\frac{{d}}{2}}}{M^{\frac{{- d}}{2}}(1+i
\varepsilon)^{-\frac{d}{2}}} (\tau_2-\tau_1)^\frac{d}{2}{V_{{\mathbb{R}^{d}}}},
\end{equation}
    where $V_{\mathbb{R}^{d}}$ is the volume of $\mathbb{R}^{d}$ space and we
    have resorted to the standard Gaussian integral
    \cite{smirnov2004evaluating}:
    \begin{equation}\label{eqGaussianInt} \int\! {{d^\delta}x} \, {e^{i\beta
      {{\vec x}^2} - 2i\vec q \cdot \vec x}} = {e^{\frac{{i\pi }}{2}\left( {1
  - \frac{\delta}{2}} \right)}}{\pi ^{\delta/2}}{\beta ^{ - \delta/2}}{e^{
  - i\frac{{{{\vec q}^2}}}{\beta }}}.  \end{equation}
    We can check this by setting $\xi  = \frac{M(1+i \varepsilon)}{{2({\tau_2}
    - {\tau_1})}}$, thus
    \begin{align*}\label{key} {I_x} &= \int\! {{d^d}{x_1} \int\! {{d^d}{x_2} \,
      {e^{i\xi {{({{\vec x}_2} - {{\vec x}_1})}^2}}}} }  = \int\! {{d^d}{x_1}
    \int\! {{d^d}{x_2} \,{e^{i\xi (\vec x_1^2 - 2\vec x_1.{{\vec x}_2} + \vec
  x_2^2)}}} } \\ &= \int\! {{d^d}{x_1} \, {e^{i\xi \vec x_1^2}} \underbrace
{\int\! {{d^d}{x_2} \, {e^{i\xi \vec x_2^2 - 2i\xi \vec x_1.{{\vec x}_2}}}}
}_{{e^{\frac{{i\pi }}{2}\left( {1 - \frac{d}{2}} \right)}}{\pi ^{d/2}}{\xi ^{
- d/2}}{e^{ - i\frac{{{{(\xi \vec x_1^{})}^2}}}{\xi }}}}}  = {e^{\frac{{i\pi
}}{2}\left( {1 - \frac{d}{2}} \right)}}{\pi ^{d/2}}{\xi ^{ - d/2}}\underbrace
{\int\! {{d^d}{x_1} \, {e^{i\xi \vec x_1^2}} \, {e^{ - i\xi \vec x_1^2}}}
}_{{V_{{\mathbb{R}^d}}}}\\ & = {e^{\frac{{i\pi }}{2}\left( {1 - \frac{d}{2}}
\right)}}{2^{d/2}}{\pi ^{d/2}}{M^{ - d/2}}(1+i
\varepsilon)^{-\frac{d}{2}}{({\tau_2} - {\tau_1})^{d/2}}{V_{{\mathbb{R}^d}}}.
\end{align*}
    The integrals over $\tau_1$ and $\tau_2$ now take the form
    \begin{equation}\label{eqTimeIntA} {I_t} = \int\limits_{ - \infty }^{
      - \epsilon }\! {d{\tau_1} \int\limits_\epsilon^\infty\!  {d{\tau_2} \,
    {{({\tau_2} - {\tau_1})}^{\frac{d}{2} - \Delta }}}
  }  = \frac{{{2^{\frac{d}{2} - \Delta  + 4}}}}{{(d - 2\Delta  + 2)(d - 2\Delta
+ 4)}}{\epsilon ^{\frac{d}{2} - \Delta  + 2}}, \end{equation}
    where the integrals are convergent only if 
    \begin{equation}\label{eqDivCondQuantumFisherA} 2+\frac{d}{2}-\Delta<0.
    \end{equation}
    Therefore, QIM in the dual CFT to Schr\"odinger spacetime yields
    \begin{equation}\label{eqCFTDualFIMA} {G_{\lambda \lambda }^{(CFT)}}
      = \frac{{\Delta {2^{d - 2\Delta  + 3}}{e^{- \frac{{i\pi }}{4}(d +2\Delta
      )}}{M^{ \Delta - \frac{d}{2}  - 1}}{V_{{\mathbb{R}^d}}}}}{{(d - 2\Delta
  + 4) \,\Gamma \left( {\Delta  - \frac{d}{2}} \right) {(1+i
  \varepsilon)^{\frac{d}{2}}}}}{\,\epsilon ^{2 + \frac{d}{2} - \Delta }}=C
  {\epsilon ^{2 + \frac{d}{2} - \Delta }}, \end{equation} where at the final
  expression we have removed the regulator $ \varepsilon$ from the correlation
  function.  \section{Computation of the bulk holographic Fisher
  metric}\label{appB}
		
    \subsection{Computation of the fields $\phi_{1,2}$} We will be integrating
    the bulk-to-boundary propagator $K(r,\vec x,\tau;{{\vec x}_1},{\tau_1})$
    over $x_1$ and $\tau_1$, when $\tau_1<\tau$. Therefore, the field $\phi_1$
    is given by\footnote{Here $\hat\phi _0(\vec x_1,\tau_1)=1$.}
		
    \begin{align}\label{key} \nonumber {\phi _1}(r,\vec x,\tau) &= \delta
      \lambda \int\! {{d^d}{x_1}\int\limits_{ -\infty }^\tau\! {d{\tau_1} \,
      K(r,\vec x,\tau;{{\vec x}_1},{\tau_1})} \, {\hat\phi _0(\vec
  x_1,\tau_1)}} \\\nonumber &= \delta\lambda\, {c_\Delta }{r^\Delta
}\int\limits_{ -\infty }^\tau\! {d{\tau_1} \, \frac{1}{{{{(\tau
- {\tau_1})}^\Delta }}}\, {e^{\frac{i\, r^2 M}{2(\tau - \tau_1) }}}} \,
{e^{\frac{i\, M\vec x^{2} }{2(\tau - {\tau_1})}}} \int\! {{d^d}{x_1}} \,
{e^{\frac{i M }{2(\tau - \tau_1)}(\vec x_1^2 - 2\vec x_1.{{\vec
x}})}}\\\nonumber &=\delta \lambda {c_\Delta }{r^\Delta }\int\limits_{-\infty
}^\tau\! d{\tau_1} \, \frac{1}{{{{(\tau - {\tau_1})}^\Delta }}}\, e^{\frac{i\,
r^2 M}{2(\tau - \tau_1)}} \, e^{\frac{i\, M \vec x^2}{2(\tau - \tau_1)}} \,
e^{\frac{ -i\,M\vec x^2}{2(\tau - \tau_1)}} \, e^{\frac{{i\pi }}{2}\left( {1
- \frac{d}{2}} \right)} {\pi ^{d/2}} \left( \frac{M}{2(\tau - \tau_1)}
\right)^{ - d/2}\\\nonumber &=\delta \lambda\, c_\Delta\,  r^{\Delta} \,2^{d/2}
M^{-d/2}\,  e^{\frac{{i\pi }}{2}\left( {1 - \frac{d}{2}} \right)}{\pi
^{d/2}}\int\limits_{ -\infty }^\tau\! {d\tau_1\, {\mkern 1mu} e^{\frac{i\,
\mu}{\tau - \tau_1}}} {(\tau - {\tau_1})^{\frac{d}{2} - \Delta }}, \end{align}
    where we have introduced the notation
    \begin{equation}\label{key} \mu  = \frac{M r^2}{2}.  \end{equation}
    Let us calculate the last integral. For this purpose, we change the
    variables to $y=\tau-\tau_1$ with boundaries
    \begin{equation}\label{key} y = \tau - {\tau_1} = \left\{ \begin{array}{l}
    0,\quad {\tau_1} \to \tau,\\ \infty ,\quad {\tau_1} \to -\infty
, \end{array} \right.  \end{equation}
    hence,
    \begin{equation}\label{key} \int\limits_0^\infty\! dy\,  e^{\frac{i\mu}{y}}
    \, y^{\frac{d}{2} - \Delta} =  \Gamma \left( \Delta  - \frac{d}{2}
  - 1 \right)  (-i\mu)^{\frac{d}{2} - \Delta  + 1}, \end{equation}
    where one has the convergence condition
    \begin{equation}\label{eqConvCond1B} \Delta >\frac{d}{2}+1,\quad M>0.
    \end{equation}
    The field $\phi_1$ now becomes
    \begin{equation}\label{key} {\phi _1}(r,\vec x,\tau) \equiv {\phi _1}(r)
    = i\delta \lambda\, e^{ - \frac{{i\pi d}}{2}} {r^{d - \Delta  + 2}}.
  \end{equation}
		
    We can do a similar computation for the filed $\phi_2$, namely
    \begin{align}\label{key} \nonumber {\phi _2}(r,\vec x,t) &=\delta \lambda\,
      c_\Delta\,  r^{\Delta} \,2^{d/2} M^{-d/2}\, {e^{\frac{{i\pi }}{2}\left(
      {1 - \frac{d}{2}} \right)}}{\pi ^{d/2}} \int\limits_{0 }^\tau\! d\tau_1\,
      e^{-\frac{{{\tilde \mu}}}{{\tau - {\tau_1}}}} {(\tau
      - {\tau_1})^{\frac{d}{2} - \Delta }}\\\nonumber &=\delta \lambda\,
      c_\Delta\,  r^{\Delta} \,2^{d/2} M^{-d/2}\, {e^{\frac{{i\pi }}{2}\left(
      {1 - \frac{d}{2}} \right)}}{\pi ^{d/2}} \int\limits_0^\tau\! d{y}\,
      e^{-\frac{{{\tilde \mu}}}{{y}}} {y^{\frac{d}{2} - \Delta }} \\ &=\delta
      \lambda\, c_\Delta\,  r^{\Delta} \,2^{d/2} M^{-d/2}\, e^{\frac{{i\pi
      }}{2}\left( {1 - \frac{d}{2}} \right)} \pi^{d/2}\,
      \tilde\mu^{\frac{d}{2}-\Delta+1}\! \int\limits_{ \tilde
      \mu/\tau}^\infty\! d{x}\, e^{-x} {x^{a-1 }} =\frac{i \delta\lambda\,
    e^{-\frac{i \pi d}{2}} r^{d-\Delta+2}}{\Gamma(a)}
    \Gamma\left(a,\frac{\mu}{t}\right)\!, \end{align}
    where at the end we have returned to real time $\tau\to -i t$, and we have
    defined the parameter \begin{equation}\label{key} \tilde\mu  = -i\frac{M
    r^2}{2}.  \end{equation}

    \subsection{Computation of the HFIM integrals}\label{appB2}
		
    In order to compute Eqs. (\ref{eqDI1}) and (\ref{eqDI2}) we have to
    consider the metric $\gamma_{ab}$, $a,b=0,\dots, d$, on the
    ($d+1$)-dimensional boundary of the Schr\"odinger background, namely
    \begin{equation}\label{key} ds_{{{(\partial Schr)}_{d + 1}}}^2 =  - {\mkern
    1mu} {L^2}\frac{{d{t^2}}}{{{r^4}}} + {L^2}\frac{{d{{\vec x}^{{\kern 1pt}
2}}}}{{{r^2}}}, \end{equation}
    where $\vec x$ is a $d$-dimensional vector. Therefore, one finds 
    \begin{equation}\label{key} \sqrt{|\gamma|}=\frac{L^{d+1}}{r^{d+2}}.
    \end{equation}
    On the other hand, only the $n_r$ component of the normal vector to the
    boundary contributes, which can be explicitly calculated from 
    \begin{equation}\label{key} 1=g_{\mu\nu}n^\mu n^\nu=g_{rr} n^r
    n^r=\frac{L^2}{r^2}(n^r)^2.  \end{equation}
    Hence $n^r=r/L$ and consequently $n_r=g_{rr} n^r=L/r$. Finally, one has
    \begin{equation}\label{key} \sqrt{|\gamma|}\, n_\mu
    g^{\mu\nu}=\sqrt{|\gamma|}\, n_r g^{rr}={L^d}\, r^{-1-d}.  \end{equation}
		
    Now, we can proceed with the computation of the bulk holographic Fisher
    information metric, which can be written by
    \begin{equation}\label{key} {G_{\lambda \lambda }^{Bulk}} = -\,
      \frac{1}{{\delta {\lambda ^2}}}\left( \frac{\delta I_1}{2} - \delta I_2
      \right) = \frac{{{L^{d }}{V_{{\mathbb{R}^d}}}}}{{2\kappa }}\mathop {\lim
    }\limits_{r \to \tilde \varepsilon } \left( {{c_0}{J_0}\, r^{d-2 \Delta +2}
  + {c_1}{J_1} + {c_2}{J_2}\, r^{d-2 \Delta +2}} \right), \end{equation}
    where the coefficients $c_i$ are given by \begin{align}\label{key} c_0
    = \frac{e^{-i \pi d}}{2} (d-\Delta+2), \quad  c_1 = \frac{M^a \, e^{-id\pi}
  }{2^{a-1}\, \Gamma ^2(a)} , \quad c_2 = -\, \frac{2 c_0}{\Gamma^2(a)}.
\end{align}
    Consequently, one has to compute the following integrals
    \begin{align}\label{key} &{J_0} = \int\limits_ {-T}^{-\epsilon}\! dt\, +\,
      \int\limits_{\epsilon}^{T}\! dt  = 2(T - \epsilon),\\ &{J_1}
      =  \int\limits_{\epsilon}^{T}\! dt \, t^{-a} \, e^{-\frac{\mu} {t}}\,
      \Gamma \left( {a,\frac{\mu }{{t}}}
      \right)=\int\limits_{1/T}^{1/\epsilon}\! dx\, x^{a-2}\, e^{-\mu x}\,
      \Gamma(a,\mu x),\\ &{J_2} =  \int\limits_{\epsilon}^{T}\! dt \,\,
    \Gamma^2\! \left( a,\frac{\mu }{t} \right).  \end{align}
		
    In order to solve $J_1$ we transform the incomplete gamma function in the
    following way
    \begin{equation} \Gamma (a,\mu x) = (a-1)(a-2)\, \Gamma (a-2,\mu x) \,+\,
    (a-1)\, \mu^{a-2}\, x^{a-2}\, e^{-\mu x} \,+\, \mu^{a-1}\, x^{a-1}\,
  e^{-\mu x}, \end{equation}
    which follows directly from
    \begin{equation}\label{key} \Gamma (a,z) = \frac{{\Gamma (a)}}{{\Gamma (a
    - n)}}\, \Gamma (a-n,z) \,+\, z^{a-1}\, e^{-z} \sum\limits_{k = 0}^{n - 1}
  {\frac{{\Gamma (a)}}{{\Gamma (a - k)}}}\, z^{ - k} \end{equation}
    for $n=2$ and we have also used $\Gamma(a)=(a-1) \Gamma(a-1)$. Hence, the
    integral $J_1$ now becomes
    \begin{align} \nonumber {J_1} &= (a-1)(a-2)
      \int\limits_{1/T}^{1/\epsilon}\! e^{ - \mu x}\, x^{a - 2}\, \Gamma
      (a-2,\mu x)\, dx  \\\nonumber &+ (a-1)\, \mu^{a-2}\!
      \int\limits_{1/T}^{1/\epsilon}\! x^{2a - 4}\, e^{ - 2\mu x}\, dx  \,+\,
      \mu^{a-1}\! \int\limits_{1/T}^{1/\epsilon}\! x^{2a - 3}\, e^{ - 2\mu x}\,
      dx \\\nonumber &= \frac{1}{\mu}(a-1)(a-2)\, T^{2-a}\, e^{-\frac{\mu
      }{T}}\, \Gamma\! \left( a-2,\frac{\mu}{T} \right) \,+\,
      \frac{1}{2\mu^{a-1}} (a-1)(a-2)^2 \, \Gamma^2 {\left( {a - 2,\frac{\mu
      }{T}} \right)}\\\nonumber &- \frac{1}{\mu}(a-1)(a-2)\, \epsilon^{2-a}\,
      e^{-\frac{\mu }{\epsilon}}\, \Gamma\! \left( a-2,\frac{\mu}{\epsilon}
      \right) \,-\, \frac{1}{2\mu^{a-1}} (a-1)(a-2)^2\, \Gamma^2 {\left( {a
    - 2,\frac{\mu }{\epsilon }} \right)}\\\nonumber &+
    \frac{4^{2-a}}{\mu^{a-1}} (a-1)(a-2)\, \Gamma\! \left(
    2a-4,\frac{2\mu}{\epsilon} \right) \,-\, \frac{4^{2 - a}}{\mu^{a-1}}
    (a-1)(a-2)\, \Gamma\! \left( 2a-4, \frac{2\mu}{T} \right)\\\nonumber &+
    \frac{2^{3-2a}}{\mu^{a-1}} (a - 1)\, \Gamma\! \left( {2a - 3,\frac{{2\mu
    }}{T}} \right) \,-\, \frac{2^{3-2a}} {\mu^{a-1}} (a - 1)\, \Gamma\! \left(
  {2a - 3,\frac{{2\mu }}{\epsilon }} \right) \\ &+ \frac{4^{1-a}}{\mu^{a-1}}\,
\Gamma\! \left( 2a-2, \frac{2\mu}{T} \right) \,-\, \frac{4^{1-a}}{\mu^{a-1}}\,
\Gamma\! \left( 2a-2, \frac{2\mu}{\epsilon} \right).  \end{align}
		
    The integral $J_2$ is more complicated, but also analytically solvable if
    we use the following power reduction formula \cite{article}:
    \begin{equation}\label{key} \Gamma^2 (a,z) = \Gamma^2 (a) \,-\,
    2\int\limits_0^{1/2}\! d\omega {\mkern 1mu} \, \,\gamma \left(
  2a,\frac{z}{\omega} \right) \omega^{a-1} (1 - \omega )^{a-1}
.  \end{equation}
    Let $\epsilon=T_1$ and $T=T_2$, hence
    \small \begin{align*} J_2 &= \int\limits_{T_1}^{T_2}\! dt\,\, \Gamma^2\!
      \left( {a,\frac{\mu }{t}} \right) = \int\limits_{T_1}^{T_2}\! dt \left[
      \Gamma^2 (a) \,-\, 2\int\limits_0^{1/2}\! d\omega \, \gamma \left(
    2a,\frac{\mu } {\omega t} \right) \omega^{a-1}  \left(1-\omega \right)^{a
- 1} \right]\\ &= \left( T_2 - T_1 \right) \Gamma ^2 (a) \,-\,
2\int\limits_0^{1/2}\! d\omega \, \omega^{a-1} (1 - \omega )^{a - 1} \,
\int\limits_{T_1}^{T_2}\! dt\, \gamma \left( 2a,\frac{\mu } {\omega t} \right)
            \\ &= \left( T_2-T_1 \right) \Gamma^2 (a) \,-\,
            2\int\limits_0^{1/2}\! d\omega \, \omega^{a-1} (1- \omega)^{a-1}
            \left( \,\int\limits_{T_1}^{T_2}\! dt\, \Gamma (2a)
            - \int\limits_{T_1}^{T_2}\! dy\, \Gamma\! \left( 2a,\frac{\mu
          }{\omega t} \right) \right) \\ &= \left( T_2-T_1 \right) \left[
      \Gamma^2 (a) - 2\Gamma (2a)\, \mathcal{B}_{1/2} (a,a) \right] \,+\,
    2\int\limits_0^{1/2}\! d\omega \, \omega^{a-1} (1- \omega)^{a - 1}
  \mathcal{J}(\omega ) .  \end{align*}
    \normalsize The inner integral is
    \begin{align}\label{key} \nonumber\mathcal{J} (\omega ) &=
      \int\limits_{T_1}^{T_2}\! dt\, \Gamma\! \left( 2a,\frac{\mu }{\omega t}
      \right)\\ &= T_2\, \Gamma\! \left( 2a,\frac{\mu }{\omega T_2} \right)
      - T_1\, \Gamma\! \left( 2a,\frac{\mu }{\omega T_1} \right)
      + \frac{\mu}{\omega} \left[ \Gamma\! \left( 2a - 1,\frac{\mu } {\omega
      T_1} \right) - \Gamma\! \left( 2a - 1,\frac{\mu }{\omega T_2} \right)
    \right], \end{align}
    thus 
    \begin{equation}\label{key} J_2 = \left( T_2 - T_1 \right)\left[ \Gamma^2
    (a) - 2\Gamma (2a)\, \mathcal{B}_{1/2} (a,a) \right] + {I_1} + {I_2}
  + {I_3} + {I_4}, \end{equation}
    where
    \begin{align} &I_1 = -\, 2T_1 \int\limits_0^{1/2}\! d\omega \, \omega^{a-1}
      (1-\omega )^{a-1}\, \Gamma\! \left( 2a,\frac{\mu }{\omega T_1} \right)
      ,\\ &I_2 = 2T_2 \int\limits_0^{1/2}\! d\omega \, \omega^{a-1} (1 - \omega
      )^{a - 1}\, \Gamma\! \left( 2a,\frac{\mu}{\omega T_2} \right) ,\\ &I_3
      = 2\mu \int\limits_0^{1/2}\! d\omega \, \omega^{a - 2} (1 - \omega )^{a
    - 1}\, \Gamma\! \left( 2a - 1,\frac{\mu }{\omega T_1 } \right) ,\\ &I_4
    =  -\, 2\mu \int\limits_0^{1/2}\! d\omega \, \omega^{a - 2} (1 - \omega
    )^{a - 1}\, \Gamma\! \left( 2a - 1,\frac{\mu }{\omega T_2 } \right)
  .  \end{align}

    The solution is the following. Let $\mathfrak{b}_{1,2}=\mu/T_{1,2}$, then
    \begin{align*}\label{key} I_{1,2} &=  \mp 2T_{1,2} \int\limits_0^{1/2}\!
      d\omega \, \omega^{a - 1} (1 - \omega )^{a - 1}\, \Gamma\! \left(
      {2a,\frac{\mathfrak b_{1,2}}{\omega }} \right) \\ &=  \mp 2T_{1,2}
      \int\limits_0^{1/2}\! d\omega \, \omega^{a - 1} (1-\omega )^{a - 1}
      {\left( \frac{ \mathfrak b_{1,2}} {\omega} \right)}^{2a} e^{
      -\frac{\mathfrak b_{1,2}}{\omega} }  \int\limits_0^\infty  \frac{e^{
    - \frac{\mathfrak b_{1,2}}{\omega}z} } {(1 + z)^{1 - 2a} } \,dz\\ &=  \mp
    2T_{1,2} \int\limits_0^\infty\! dz\, (1 + z)^{2a-1}\, \mathfrak
    b_{1,2}^{2a} \int\limits_0^{1/2}\! d\omega\, \frac{1}{\omega ^2}   \left(
    {\frac{1}{\omega } - 1} \right)^{a - 1} e^{-\frac{\mathfrak b_{1,2}}{\omega
  }(1 + z)}\, \\ &=  \mp 2T_{1,2} \int\limits_0^\infty\! dz\, (1 + z)^{2a-1}\,
  \mathfrak b_{1,2}^{2a} \int\limits_0^{1/2}\! d\left( \frac{ -1}{\omega}
  \right)\! \left( \frac{1}{\omega} - 1 \right)^{a-1} e^{- \frac{\mathfrak
b_{1,2}}{\omega }(1 + z)} \\ &= \left\{ {x = \frac{1}{\omega } = \left\{
\begin{array}{l} 2,\quad\,\,\, \omega  = 1/2,\\ \infty ,\quad \omega
= 0 \end{array} \right.\quad } \right\} = \\ &=  \mp 2T_{1,2}
\int\limits_0^\infty\!  dz\, (1 + z)^{2a - 1}\, \mathfrak b_{1,2}^{2a}
\int\limits_2^\infty\!  dx\, \left(x - 1\right)^{a - 1} e^{ - \mathfrak b_{1,2}
(1 + z)x} \\ &=  \mp 2T_{1,2} \int\limits_0^\infty\!  dz\, (1 + z)^{a - 1}\,
\mathfrak b_{1,2}^a\, e^{ - \mathfrak b_{1,2}(1 + z)}\, \Gamma\! \left(
a,\mathfrak b_{1,2}(1 + z) \right)\\ &=  \pm T_{1,2} \int\limits_0^\infty\!
dz\, \frac{d}{dz} \Gamma^2\! \left( a,\mathfrak b_{1,2}(1 + z) \right) =  \pm
T_{1,2}\, \Gamma ^2 \left( a,\mathfrak b_{1,2}(1 + z) \right)\left|
{\begin{array}{*{20}{l}} {^{z \to \infty }}\\ {_{z \to 0}} \end{array}} \right.
=  \mp T_{1,2}\, \Gamma^2\left( {a,\mathfrak b_{1,2}} \right).  \end{align*}
    Therefore, one finds
    \begin{equation}\label{key} {I_{1,2}} =  \mp T_{1,2}\, \Gamma ^2\! \left(
    a,\frac{\mu } {T_{1,2}} \right).  \end{equation}
    In the previous computations we have used the following integral
    representation of the incomplete gamma function
    \begin{equation}\label{key} \Gamma (a,t) = t^a\, e^{-t}
    \int\limits_0^\infty  \frac{ e^{ -zt}} {(1 + z)^{1-a} }dz.  \end{equation}
		
    With similar calculations for $I_{3,4}$, one finds
    \begin{align*} I_{3,4} &=  \pm 2\mu \int\limits_0^{1/2}\! d\omega\,
      \omega^{a-2}\, (1-\omega )^{a-1} \,\Gamma\! \left( 2a-1,\frac{\mathfrak
      b_{1,2}} {\omega} \right) \\ &=  \pm 2\mu \int\limits_0^{1/2}\! d\omega\,
      \omega^{a-2}\, (1-\omega )^{a-1} \left( \frac{\mathfrak b_{1,2}}{\omega}
      \right)^{2a-1} e^{ -\frac{\mathfrak b_{1,2}}{\omega}}
      \int\limits_0^\infty  \frac{e^{ - \frac{\mathfrak b_{1,2}}{\omega}z}} {(1
      + z)^{2 - 2a} } \,dz\\ &=  \pm 2\mu \int\limits_0^\infty\! dz\,
      (1+z)^{2a-2}\, \mathfrak b_{1,2}^{2a-1} \int\limits_0^{1/2}\! d\omega\,
      \frac{1}{\omega^2} \left( \frac{1}{\omega} - 1 \right)^{a - 1} e^{
      -\frac{\mathfrak b_{1,2}}{\omega }(1 + z)} \, \\ &=  \pm 2\mu
      \int\limits_0^\infty\! dz\, (1+z)^{2a-2}\, \mathfrak b_{1,2}^{2a-1}
      \int\limits_0^{1/2}\! d\left( \frac{-1}{\omega} \right)\! \left(
      \frac{1}{\omega} -1 \right)^{a-1} e^{ -\frac{\mathfrak b_{1,2}}
    {\omega}(1+z)} \\ &= \left\{ {x = \frac{1}{\omega } = \left\{
  \begin{array}{l} 2,\quad\,\,\,  \omega  = 1/2,\\ \infty ,\quad \omega
  = 0 \end{array} \right.\quad } \right\} = \\ &=  \pm 2\mu
  \int\limits_0^\infty\! dz\, (1+z)^{2a-2}\, \mathfrak b_{1,2}^{2a-1}
  \int\limits_2^\infty\!  dx\, \left( x-1 \right)^{a-1} e^{ -\mathfrak
  b_{1,2}(1+z)x} \\ &=  \pm 2\mu \int\limits_0^\infty\! dz\, (1+z)^{a-2}\,
\mathfrak b_{1,2}^{a-1}\, e^{ - \mathfrak b_{1,2}(1 + z)}\, \Gamma\! \left(
a,\mathfrak b_{1,2} (1+z) \right).  \end{align*}
    Now we use 
    \begin{equation}\label{key} \Gamma (a,\mathfrak b(z + 1)) = {e^{
      - \mathfrak b(z + 1)}}{\mathfrak b^{a - 1}}{(z + 1)^{a - 1}}
      - \frac{1}{2}(a - 1){(z + 1)^{2 - a}}{\mathfrak b^{1 - a}}{e^{\mathfrak
      b(z + 1)}}\frac{d}{{dz}}\Gamma^2 (a - 1,\mathfrak b(z + 1)),
    \end{equation}
    thus
    \begin{align*} I_{3,4} &=  \mp (a-1)\mu \int\limits_0^\infty\! dz\,
      \frac{d}{dz} \Gamma^2 (a-1,\mathfrak b_{1,2} (z + 1))  \,\pm\,  2\mu
      \mathfrak b_{1,2}^{2a-2} \int\limits_0^\infty\!  dz\, (z + 1)^{2a-3}\,
      e^{ -2\mathfrak b_{1,2} (z + 1)}\\ &=  \mp (a - 1)\mu\, \Gamma^2\!
    \left(a-1, \mathfrak b_{1,2}(z+1)\right) \left| \begin{array}{l} ^{z \to
  \infty }\\ _{z \to 0} \end{array} \right. \pm\, 2^{3-2a}\mu\, \Gamma
  (2a-2,2\mathfrak b_{1,2})\\ &=  \pm (a - 1)\mu\, \Gamma^2 (a - 1,\mathfrak
b_{1,2})  \,\pm\,  2^{3 - 2a}\mu\, \Gamma (2a - 2,2\mathfrak b_{1,2}).
\end{align*}
    Therefore, one has
    \begin{equation}\label{key} I_{3,4} =  \pm \mu \left[ (a-1)\, \Gamma^2\!
    \left( a-1,\frac{\mu } {T_{1,2}} \right) + 2^{3-2a}\, \Gamma\! \left(
2a-2,\frac{2\mu} {T_{1,2}} \right) \right].  \end{equation}
    Putting everything together we get the complete expression for $J_2$,
    namely
    \begin{align} \nonumber J_2 &= (T-\epsilon )\left(  \Gamma^2 (a) - 2\Gamma
      (2a)\, \mathcal{B}_{1/2}(a,a) \right) \,+\, T\,\Gamma^2\! \left(
    a,\frac{\mu}{T} \right) - \epsilon\, \Gamma^2\! \left(
  a,\frac{\mu}{\epsilon} \right) \\\nonumber &+ \mu \left( (a-1)\, \Gamma^2\!
\left( a-1,\frac{\mu}{\epsilon} \right) + 2^{3-2a}\, \Gamma\! \left(
2a-2,\frac{2\mu}{\epsilon } \right) \right) \\ &- \mu \left( (a - 1)\,
\Gamma^2\! \left( a-1,\frac{\mu }{T} \right) + 2^{3-2a}\, \Gamma\! \left( 2a-2,
\frac{2\mu}{T} \right) \right).  \end{align}
		
    The final expression for the holographic Fisher metric yields
    \begin{align}\label{eqFisherBulkB} \nonumber {G_{\lambda \lambda }^{Bulk}}
    &= \frac{a_0}{\mu^{a}} (T - \epsilon ) \,+\, a_1 \frac{T^{2 - a} e^{
    -\frac{\mu}{T}}}{\mu} \,  \Gamma\! \left( a-2,\frac{\mu}{T} \right) \,+\,
    \frac{a_2}{\mu^{a-1}}\, \Gamma^2\! \left( a-2,\frac{\mu}{T}
    \right)\\\nonumber &+ \frac{a_3}{\mu^{a-1}}\, \Gamma\! \left( 2a
  - 4,\frac{2\mu}{T} \right) \,+\, \frac{a_4}{\mu^{a-1}}\, \Gamma\! \left(
2a-2,\frac{2\mu}{T} \right) \,+\, \frac{a_5}{\mu^{a-1}}\, \Gamma\! \left(
2a-3,\frac{2\mu}{T} \right)\\\nonumber & + a_6 \frac{T}{\mu^a}\, \Gamma^2\!
\left( a,\frac{\mu }{T} \right) \,+\, \frac{a_7}{\mu^{a-1}}\, \Gamma^2\! \left(
a-1,\frac{\mu }{T} \right)\\\nonumber &+ b_1 \frac{\epsilon^{2-a}}{\mu}\, e^{
-\frac{\mu}{\epsilon}}\, \Gamma\! \left( a-2,\frac{\mu}{\epsilon} \right) \,+\,
\frac{b_2}{\mu^{a-1}}\, \Gamma^2\! \left( a-2,\frac{\mu }{\epsilon}
\right)\\\nonumber & + \frac{b_3}{\mu^{a-1}}\, \Gamma\! \left( 2a-4,\frac{2\mu
}{\epsilon} \right) \,+\, \frac{b_4}{\mu^{a-1}}\, \Gamma\! \left(
2a-2,\frac{2\mu}{\epsilon } \right) \,+\, \frac{b_5}{\mu^{a-1}}\, \Gamma\!
\left( 2a-3,\frac{2\mu} {\epsilon} \right)\\ & + b_6 \frac{\epsilon}
{\mu^{a}}\, \Gamma^2\! \left( a,\frac{\mu}{\epsilon} \right) \,+\,
\frac{b_7}{\mu^{a-1}}\, \Gamma^2\! \left( a-1,\frac{\mu}{\epsilon} \right),
\end{align}
    with coefficients $a_i=-b_i$, $i=1,\dots, 7$, where
    \begin{align} \nonumber &{a_0} = \frac{{{L^{d
      }}{V_{{\mathbb{R}^d}}}}}{2^{a+1}\kappa } M^a \left[ {{2c_0} + {c_2}\left(
        {{\Gamma ^2}\left( a \right) - 2\Gamma (2a)\, \mathcal{B}_{1/2} (a,a)}
        \right)} \right],\\\nonumber &{a_1} =  - {b_1} = \frac{{{L^{d
  }}{V_{{\mathbb{R}^d}}}}}{{2\kappa }}{c_1}(a - 1)(a - 2),\\\nonumber &{a_2}
  =  - {b_2} = \frac{{{L^{d }}{V_{{\mathbb{R}^d}}}}}{{4\kappa }}{c_1}{(a
  - 1)}(a - 2)^2,\\\nonumber &{a_3} =  - {b_3} =  - \frac{{{L^{d
}}{V_{{\mathbb{R}^d}}}}}{{2^{2a-3}\kappa }} c_1(a - 1)(a - 2),\\\nonumber
&{a_4} =  - {b_4} = \frac{{{L^{d }}{V_{{\mathbb{R}^d}}}}}{{2^{3a-1}\kappa }}
         \left(2^a {c_1}-2 {c_2} M^a\right),\\\nonumber &{a_5} =  - {b_5}
         = \frac{{{L^{d }}{V_{{\mathbb{R}^d}}}}}{{2^{2a-2}\kappa }} c_1 (a
         - 1),\\\nonumber &{a_6} =  - {b_6} = \frac{{{L^{d
       }}{V_{{\mathbb{R}^d}}}}}{{2^{a+1}\kappa }} M^{a} c_2,\\ &{a_7}
     =  - {b_7} =  - \frac{{{L^{d }}{V_{{\mathbb{R}^d}}}}}{{2^{a+1}\kappa }}
   M^{a} c_2(a - 1), \end{align} where $\mathcal{B}$ is the incomplete beta
   function.  \end{appendix}

	
  \bibliographystyle{utphys} \bibliography{TS-refs}
	
\end{document}